\documentclass[a4paper,11pt]{article}

\usepackage[utf8]{inputenc}
\usepackage{geometry}
\usepackage{graphicx}
\usepackage{color}
\usepackage{amsmath}
\usepackage{amssymb}
\usepackage{lineno}
\usepackage{enumitem}
\usepackage{array}
\usepackage{hyperref}
\usepackage{upgreek}
\usepackage{siunitx}
\usepackage[normalem]{ulem}
\usepackage{subcaption}
\usepackage{float}
\usepackage{titlesec}
\usepackage{booktabs}
\usepackage{natbib}
\usepackage{physics}

\defcitealias{Leib_Wundrow_Goldstein_1999}{LWG99}
\defcitealias{Ricco_Wu_2007}{RW07}

\def\average#1{\left\langle{#1}\right\rangle}
\def\p{\partial}
\def\abs#1{\left\vert{#1}\right\vert}
\def\Or#1{\mathcal{O}\left({#1}\right)}
\newcolumntype{P}[1]{>{\centering\arraybackslash}p{#1}}
\def\xbar{\overline{x}}
\def\utd{\overline{u}}
\def\vtd{\overline{v}}
\def\wtd{\overline{w}}
\def\tautd{\overline{\tau}}
\def\ptd{\overline{p}}

\DeclareRobustCommand{\VAN}[3]{#2} 

\title{\textbf{Receptivity of compressible boundary layers \\ over porous surfaces}}

\author{
Pierre Ricco and Ludovico Foss\`{a} \\
{\itshape Department of Mechanical Engineering, University of Sheffield,} \\
{\itshape Sheffield S1 3JD, United Kingdom}
}

\begin{document}

\DeclareRobustCommand{\VAN}[3]{#3}

\maketitle

\begin{center}
    \textcolor{blue}{\textbf{Accepted for publication in Physical Review Fluids}}
\end{center}

\begin{abstract}
Supersonic pre-transitional boundary layers flowing over porous flat and concave surfaces are studied using numerical and asymptotic methods. The porous wall is composed of thin equally-spaced cylindrical microcavities.
The flow is perturbed by small-amplitude, free-stream vortical disturbances of the convected gust type. From the proximity of the leading edge, these external agents generate the compressible Klebanoff modes, i.e. low-frequency disturbances of the kinematic and thermal kind that grow algebraically downstream. For Klebanoff modes with a spanwise wavelength comparable with the boundary-layer thickness, the porous surface has a negligible effect on their growth. When the spanwise wavelength is instead larger than the boundary-layer thickness, these disturbances are effectively attenuated by the porous surface.
For a specified set of frequency and wavelengths, the Klebanoff modes evolve into oblique Tollmien-Schlichting waves through a leading-edge-adjustment receptivity mechanism. The wavenumber of these waves is only slightly modified over the porous surface, while the growth rate increases, thus confirming previous experimental results. An asymptotic analysis based on the triple-deck theory confirms these numerical findings. When the wall is concave, the amplitude of the Klebanoff modes is enhanced by the wall curvature and is attenuated by the wall porosity during the initial development. 
\end{abstract}

\section{Introduction}
\label{sec:intro}
Passive control methods aiming to delay laminar-to-turbulent transition in high speed wall flows have been the subject of several studies of numerical, experimental and theoretical nature. The development of new flow control strategies is critical to several aerospace applications ranging from the heat-transfer management on the surface of atmospheric reentry vehicles and in supersonic transport systems \citep{martin-etal-2012,Schmisseur_2015} to the design of hypersonic quiet tunnels where the level of noise contamination has to be reduced to a minimum \citep{Schneider_2008}. 

Although the inviscid second mode of instability is predominant in high-Mach number boundary layer flows \citep{Mack_1984}, the presence of both the relatively low-frequency first-mode stability disturbances and the laminar streaks triggered by free-stream vorticity has been documented in wind-tunnel experiments. \cite{Munoz_Heitmann_Radespiel_2014} observed streaky structures in the cross flow over a cone in a Ludwieg tube at Mach 6. In the experiments of \cite{Hofferth_Humble_Floryan_Saric_2013} and \cite{Borg_Kimmel_Hofferth_Bowersox_2015}, the energy content measured at low frequency was comparable with the energy peak associated with the second instability mode at higher frequency for all the unit Reynolds numbers considered. \cite{Graziosi_Brown_2002} also measured large low-frequency disturbances in a pre-transitional Mach-3 boundary layer exposed to vortical and acoustic free-stream fluctuations.

The second mode of instability is effectively attenuated by passive porous coatings. Fedorov and co-workers first showed, in their linear stability and experimental analyses \citep{Malmuth_Fedorov_Shalaev_Cole_Hites_Williams_Khokhlov_1998,Fedorov_Malmuth_Rasheed_Hornung_2001,Rasheed_Hornung_Fedorov_Malmuth_2002,Fomin_Fedorov_Shiplyuk_Maslov_Burov_Malmuth_2002,Maslov_2003}, how the presence of a porous surface could result in an attenuation of the acoustic disturbances that propagate within the boundary layer at the expense of a slight enhancement of the Tollmien-Schlichting waves. 

This type of porous coatings has the advantage of interacting with small-amplitude disturbances without affecting the laminar base flow. These early analyses paved the way to further studies on ultrasonically absorptive coatings \citep{Fedorov_Kozlov_Shiplyuk_Maslov_Malmuth_2006,Wartemann_Ludeke_Sandham_2012,Wartemann_Wagner_Giese_Eggers_Hannemann_2014} and have been extended to include non-regular geometries, acoustic scattering effects, and coupling mechanisms between adjacent pores \citep{Zhao_Liu_Wen_Zhu_Cheng_2018,Zhao_Zhang_Wen_2020,Gui_Wang_Zhao_Zhao_Wu_2022}. Other researchers have studied surfaces with two-dimensional, equally-spaced grooves of constant width \citep{Bres_Inkman_Colonius_Fedorov_2013} and porous surfaces with non-regular microstructures \citep{sousa-etal-2019}. 
\citet{Egorov_Fedorov_Soudakov_2008} investigated the effect of a porous layer on the receptivity of a boundary layer to free-stream disturbances of the acoustic type. In his review paper, \citet{Fedorov_2011} advocated further study on porous coatings for the control of boundary-layer receptivity and transient growth.

In the words of \citet{Morkovin_1969_martin_marietta_tech_report}, the term receptivity refers to the process of internalization of the free-stream disturbances in the boundary layer, their subsequent downstream evolution and the excitation of unstable disturbances. When the disturbance amplitude is relatively high, the early stage of transition in flat-plate boundary layers is dominated by the algebraic growth of externally forced perturbations rather than the exponential amplification of normal modes. The streamwise-elongated fluid structures of nearly constant spanwise wavelength, often referred to as streaks or Klebanoff modes \citep{Kendall_1985,Kendall_1990}, reach a saturation level and break down to turbulence through a secondary-instability mechanism \citep{Ricco_Luo_Wu_2011}. Their evolution was the subject of experiments performed in incompressible boundary layers \citep{Westin_Boiko_Klingmann_Kozlov_Alfredsson_1994,Matsubara_Alfredsson_2001,Fransson_Matsubara_Alfredsson_2005}. A mathematical description of the incompressible Klebanoff modes was developed by Goldstein and co-workers \citep{Leib_Wundrow_Goldstein_1999} (hereafter referred to as \citetalias{Leib_Wundrow_Goldstein_1999}). Through an asymptotic approach, they unraveled the physical interaction between the disturbances in the free stream and the boundary layer. At downstream locations where the boundary-layer thickness is comparable to the spanwise wavelength of the disturbances, the spanwise diffusion is no longer negligible and the disturbance is described by the unsteady boundary-region equations. The mathematical formulation hinges on the assumption of streamwise-elongated structures, which results in negligible streamwise diffusion and streamwise pressure gradient. The differential problem that arises is of parabolic nature, and thus suitable to a downstream-marching treatment. The computational cost is considerably lower than that required by the numerical solution of the complete Navier-Stokes equations.  Goldstein's theory is based on the precise specification of the initial and boundary conditions and accounts for the effect of the continuous outer forcing on the growth of the Klebanoff modes as the flow evolves downstream. For a review of the theory, the reader is referred to \cite{Ricco_Walsh_Brighenti_McEligot_2016}. 

Compressible laminar boundary layers are receptive to free-stream disturbances of the vortical, acoustic, and entropic type \citep{Kovasznay_1953}. The early stages of transition in compressible boundary layers are thus considerably more complex than in the incompressible regime. The rather scarce experimental literature has mainly focused on the role of the acoustic disturbances in supersonic quiet tunnels \citep{Kendall_1975,Graziosi_Brown_2002}. However, the presence of free-stream vorticity is relevant, as all types of disturbances are present downstream of a shock wave \citep{McKenzie_Westphal_1968} and exist in supersonic and hypersonic wind tunnels \citep{Schneider_2008}. 

The linear incompressible theory of \citetalias{Leib_Wundrow_Goldstein_1999} was extended to the compressible case \citep{Ricco_Wu_2007} (\citetalias{Ricco_Wu_2007}), to the nonlinear incompressible case \citep{Ricco_Luo_Wu_2011} and to the nonlinear compressible case \citep{Marensi_Ricco_Wu_2017}. The linear theory well describes the initial growth of the Klebanoff modes, their amplitude still being small and the intermodal coupling negligible. The contribution of each monochromatic mode can be studied separately in this case. The nonlinear theory instead applies when the amplitude of the disturbance flow is comparable with that of the base flow. The combined effect of a continuous free-stream spectrum of vortical disturbances and nonlinearity within the boundary layer was considered by \cite{Zhang_Zaki_Sherwin_Wu_2011}. \cite{Wu_Dong_2016} included the contribution of short-wavelength free-stream disturbances in incompressible and compressible boundary layers. The spanwise and wall-normal wavelengths were comparable with the boundary-layer thickness, that is, they were much shorter than those considered by \citetalias{Leib_Wundrow_Goldstein_1999}.

Free-stream vorticity has also been indicated as a critical cause for the generation and growth of unsteady counter-rotating G\"{o}rtler vortices \citep{Wu_Zhao_Luo_2011,Viaro_Ricco_2018,Viaro_Ricco_2019a,Viaro_Ricco_2019b}. A laminar boundary layer over a concave wall is subject to an inviscid instability caused by the imbalance between the centrifugal force and the radial pressure gradient. Free-stream vorticity triggers the onset of Klebanoff modes near the leading edge, which, because of the wall curvature, evolve downstream into G\"{o}rtler vortices, as shown by \cite{Wu_Zhao_Luo_2011}, \cite{Xu_Zhang_Wu_2017} and \cite{Xu_Liu_Wu_2020}. 

Motivated by \citet{Egorov_Fedorov_Soudakov_2008} and \citet{Fedorov_2011}, we study the effect of porous surfaces on the receptivity of supersonic boundary layers excited by free-stream vortical disturbances and, in particular, on the generation and evolution of compressible Klebanoff modes and highly-oblique Tollmien-Schlichting waves (TS) over these porous surfaces. We adopt the porous-layer model first utilized by \citet{Fedorov_Malmuth_Rasheed_Hornung_2001}, which is characterized by a regular microstructure of thin, uniformly-spaced cylindrical pores. To our knowledge, it is the first time that porous surfaces are utilized to control Klebanoff modes over flat and concave porous surfaces. 
The mathematical framework is discussed in \S \ref{sec:mathematical}. The Klebanoff modes are studied in \S\ref{sec:k-modes} and the receptivity and exponential growth of the TS waves are investigated in \S\ref{sec:TS}. The combined effect of wall porosity and curvature is the subject of \S\ref{sec:gortler}. Conclusions are presented in \S\ref{sec:conclusions}.

\section{Mathematical formulation}
\label{sec:mathematical}
\begin{figure}
    \centering
    {\includegraphics[width=0.8\linewidth]{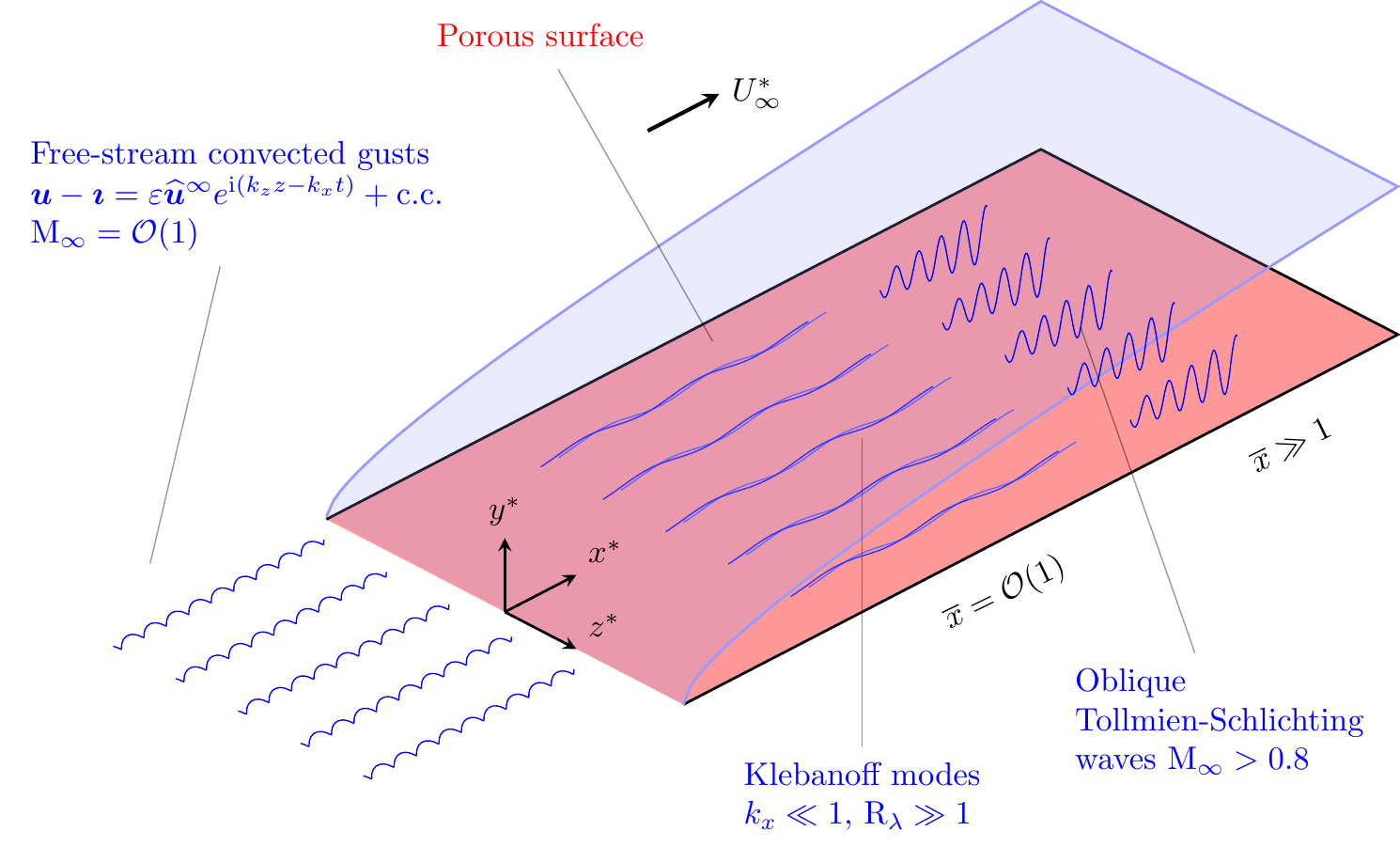} \\
    \includegraphics[width=0.8\linewidth]{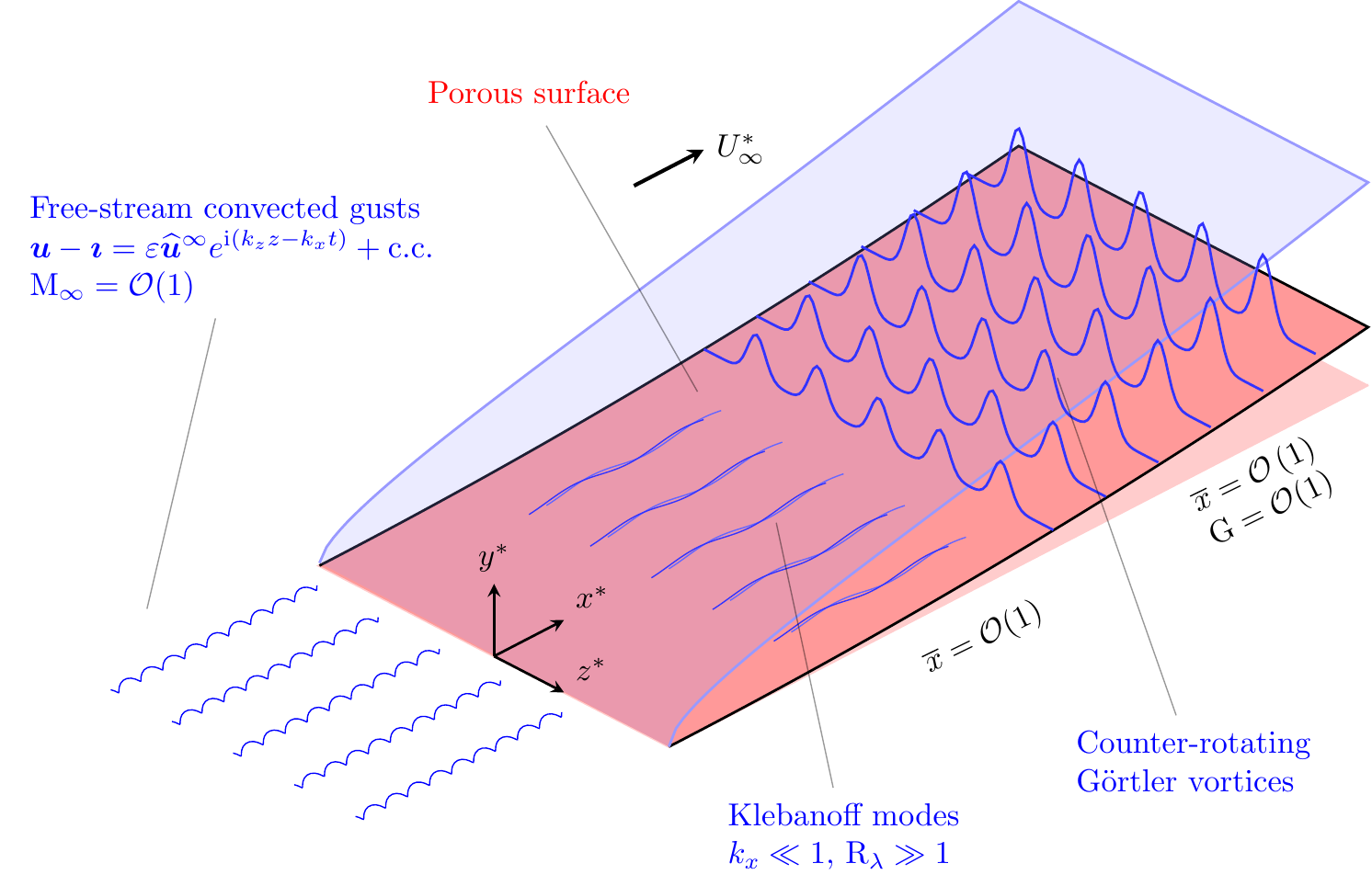}}
    \caption{Schematics of the physical domains. Flat-plate system (top) and concave-surface system (bottom).}
    \label{fig:physical_domain}
\end{figure}

A supersonic uniform air flow with free-stream velocity $U_\infty^\ast$ and static temperature $T_\infty^\ast$ past an infinitely-thin plate is considered. The flow is described in a Cartesian frame of reference, where $x^\ast$, $y^\ast$ and $z^\ast$ define the streamwise, wall-normal and spanwise coordinates, respectively. The leading edge of the plates is located at $x^\ast=y^\ast=0$. The Mach number is $\mathrm{M}_\infty\equiv U_\infty^\ast/c_\infty^\ast$, where $c_\infty^\ast=\sqrt{\gamma\mathcal{R}^\ast T_\infty^\ast}$ is the speed of sound in the free stream, $\gamma=1.4$ is the heat capacity ratio, and $\mathcal{R}^\ast=287.05\,\si{\joule\per\kilo\gram\per\kelvin}$ is the specific gas constant of air. All dimensional quantities are denoted by the superscript $^\ast$. Schematics of the physical domains are shown in figure \ref{fig:physical_domain}. Sketch a) depicts the flat-wall system where Klebanoff modes turn into TS waves and sketch b) represents the concave-wall system where Klebanoff modes turn into G\"ortler vortices. The steady compressible laminar boundary layer forming over the plate is referred to as the base flow \citep{Stewartson_1964}. The free stream is perturbed by small-amplitude, homogeneous disturbances of the convected gust type, i.e. vortical perturbations which are purely advected by the free-stream base flow. The spatial coordinates and all the boundary-layer lengths and wavenumbers are scaled by the spanwise wavelength of the gust, $\lambda_z^\ast$. The time is scaled by $\lambda_z^\ast/U_\infty^\ast$. The velocity components, the density, the viscosity and the temperature are normalized by their free-stream values and the pressure is scaled by $\rho_\infty^\ast U_\infty^{\ast2}$, where $\rho^\ast_\infty$ is the density of the fluid in the free stream.

The focus of the present work is on the early-stage growth of the laminar streaks in a pre-transitional boundary layer. As the amplitude of the perturbations is assumed small, we perform a linear analysis that supports single monochromatic disturbances as the coupling between different modes and secondary instability effects can only be captured in the nonlinear case \citep{Zhang_Zaki_Sherwin_Wu_2011}. Albeit idealized, this assumption permits to elucidate important aspects of the receptivity and early-stage growth of the boundary layer streaks \citep{Wu_Zhao_Luo_2011}. Moreover, we assume that the perturbations are of low frequency because it is well known that low-frequency, free-stream vortical disturbances are the most likely to generate streamwise-elongated structures in the boundary layer. These structures include the laminar streaks over flat plates \citep{Kendall_1985,Kendall_1990,Westin_Boiko_Klingmann_Kozlov_Alfredsson_1994,Bertolotti_1997,Fransson_Matsubara_Alfredsson_2005}, and G\"ortler vortices on concave surfaces \citep{Wu_Zhao_Luo_2011,Boiko_Ivanov_Kachanov_Mischenko_Nechepurenko_2017,Viaro_Ricco_2019a}. The spectra of streaks measured by \cite{Matsubara_Alfredsson_2001} (figure 9b therein) showed a higher energy content at low frequency and a lower energy content at high frequency compared to the free stream. 

The small-amplitude, non-interacting perturbations in the free-stream are modeled by a single monochromatic perturbation of the gust type,
\begin{equation}
\label{free-stream_perturbation}
    \boldsymbol{u}-\boldsymbol{\imath}=\varepsilon\widehat{\boldsymbol{u}}^\infty e^{\mathrm{i}\left(k_zz-k_xt\right)} + \mathrm{c.c.},
\end{equation}
where $\boldsymbol{u}$ is the free-stream velocity vector, $\boldsymbol{\imath}$ is the streamwise unit vector, $\varepsilon\ll1$ indicates the amplitude of the gust, $\widehat{\boldsymbol{u}}^\infty=\lbrace\widehat{u}^\infty, \widehat{v}^\infty, \widehat{w}^\infty\rbrace=\mathcal{O}(1)$, and c.c. its complex conjugate. The gust is characterized by a large wavelength ratio $\lambda_x^\ast/\lambda_z^\ast\gg 1$ and a small frequency $k_x=\omega^\ast\lambda_z^\ast/U_\infty^\ast \ll 1$, where $\omega^\ast$ is the angular frequency. A Reynolds number $\mathrm{R}_\lambda\equiv U_\infty^\ast\lambda_z^\ast/\nu_\infty^\ast\gg1$ is defined, where $\nu^\ast_\infty$ is the kinematic viscosity of the fluid in the free stream. 
We investigate downstream locations at which the base-flow boundary-layer thickness is $\delta=\left(2x/\mathrm{R}_\lambda\right)^{1/2}=\mathcal{O}(1)$ and the spanwise and wall-normal diffusions are comparable. A distinguished scaling $k_x=\mathcal{O}(R_\lambda^{-1})$ emerges \citepalias{Leib_Wundrow_Goldstein_1999}, as the boundary-layer disturbances evolve downstream on a length scale comparable with the gust streamwise wavelength. The disparity between the spanwise and streamwise scales results in $\mathcal{O}(\varepsilon)$ free-stream fluctuations generating $\mathcal{O}(\varepsilon/k_x)$ streamwise velocity disturbances within the boundary layer. The mass, momentum, and energy balances of the boundary-layer disturbances are described by the compressible unsteady boundary-region equations \citepalias{Ricco_Wu_2007}. The small disturbance amplitude relative to that of the base flow allows for their linearization, i.e. for $\varepsilon/k_x\ll1$ or, equivalently, $\varepsilon \mathrm{R}_\lambda\ll1$. Thorough discussions of these scaling relationships are found in \citetalias{Leib_Wundrow_Goldstein_1999}, \citetalias{Ricco_Wu_2007} and the references therein. The linearization results in a one-way coupling between the base flow and the superposed disturbances. The analysis of the nonlinear effects, which come into play when $\varepsilon\mathrm{R}_\lambda=\order{1}$, is beyond the scope of the present study. The influence of nonlinearity on the growth of laminar streaks and G\"{o}rtler vortices, where the coupling is two-way as the streaks generated within the boundary layer also affect the free-stream disturbances, has been studied by \cite{Ricco_Luo_Wu_2011,Xu_Zhang_Wu_2017,Marensi_Ricco_Wu_2017,marensi-ricco-2017} and \cite{Xu_Liu_Wu_2020}.

\subsection{The laminar base flow}
\label{sec:steady}
The steady compressible boundary-layer equations are cast into a more compact form by applying the Dorodnitsyn-Howarth coordinate transformation \citep{Stewartson_1964},

\begin{equation}
\label{eq:big_y}
\overline{Y}(x,y) \equiv \int_{0}^{y} \rho(x, \breve{y}) \ \mathrm{d} \breve{y}.
\end{equation}
In the absence of a streamwise pressure gradient, a similarity solution exists and a wall-normal similarity variable  

\begin{equation}
\label{eq:similarity_variable} 
\eta \equiv \overline{Y}\left(\frac{\mathrm{R}_\lambda}{2 x}\right)^{1/2}
\end{equation}
is defined. The streamwise velocity, the wall-normal velocity and the temperature of the base flow are

\begin{equation}
\label{base_flow_solutions}
\begin{array}{ccc}
U = F^\prime(\eta),\quad &
V =( 2 x \mathrm{R}_\lambda)^{-1/2}\left( \eta_c T F^\prime - T F \right),\quad &
T = T(\eta),
\end{array}
\end{equation}
where the prime denotes differentiation with respect to $\eta$ and 

\begin{equation}
    \eta_c \equiv \frac{1}{T} \int_0^\eta T\left(\breve{\eta}\right) \mathrm{d}\breve{\eta}.
\end{equation}
The base-flow solution \eqref{base_flow_solutions} satisfies the coupled streamwise momentum and energy balance equations

\begin{subequations}
\label{compressible_Blasius}
\begin{align}
\label{eq:Blasius_momentum}
\left[\left(\mu/T\right) F^{\prime\prime}\right]^\prime + FF^{\prime\prime} &
= 0 , \\ 
\mathrm{Pr}^{-1}\left[(\mu/T) T^\prime\right]^\prime + FT^\prime +(\gamma -1)
\mathrm{M}_\infty^2 (\mu/T) \left(F^{\prime\prime}\right)^2 &= 0, 
\label{eq:Blasius_energy}
\end{align}
\end{subequations}
subject to the boundary conditions
\begin{equation}\label{base_flow_boundary_conditions}
    \begin{array}{ccc}
        F\left(0\right) =0, & F^\prime\left(0\right)=0, &F^\prime\left(\infty\right) \to1 \\
        T(0) =T_w, & T\left(\infty\right)\to 1. &
    \end{array}
\end{equation}
The Prandtl number is $\mathrm{Pr}=0.7$. The dynamic viscosity has a power-law dependence on the temperature \citep{Cebeci_2002},
\begin{equation}
\mu  = T^n \quad \mbox{with}\quad n = 0.76.
\end{equation}
This relation is preferred to the Chapman law ($n = 1$) as a more accurate model in the supersonic regime \citep{Stewartson_1964}. Appendix \ref{app:validation} presents a validation study of the computation of the laminar base flow.

\subsection{The unsteady disturbance flow}
\label{sec:disturbance}
The boundary-layer flow is decomposed as the sum of the base flow and the small-amplitude perturbation flow,
\begin{equation}
\left\lbrace u,\, v,\, w,\, \tau,\, p\right\rbrace 
= \left\lbrace U,\, V,\, 0,\, T,\, -1/2\right\rbrace 
+ \varepsilon \left\lbrace \widetilde u, \widetilde v, \widetilde w, \widetilde \tau, \widetilde p \right\rbrace e^{\mathrm{i} \left(k_z z - k_x t\right)} + \mathrm{c.c.}, 
\end{equation}
where
\begin{equation}
\label{eq:disturbance}
\left\lbrace \widetilde{u}, \widetilde{v}, \widetilde{w}, \widetilde{\tau}, \widetilde{p} \right\rbrace
= \left\lbrace \overline{u}_0,\, \left(\frac{2\overline{x}k_x}{\mathrm{R}_\lambda}\right)^{1/2} \overline{v}_0 ,\, \overline{w}_0 ,\, \overline{\tau}_0 ,\, \left(\frac{k_x}{\mathrm{R}_\lambda}\right)^{1/2}\overline{p}_0 \right\rbrace \left(\overline{x},\eta\right).
\end{equation}
The streamwise coordinate is scaled by the gust streamwise wavenumber $k_x^\ast=2\pi/\lambda_x^\ast$, i.e. $\overline{x}=k_x x=2\pi x^\ast/\lambda^\ast_x=\mathcal{O}(1)$, where $\lambda_x^\ast$ is the gust streamwise wavelength.

Following \citet{Gulyaev_Kozlov_Kuzenetsov_Mineev_Sekundov_1989,Choudhari_1996} and \citetalias{Leib_Wundrow_Goldstein_1999}, the solution is expanded as a weighted sum of the two-dimensional $\left\lbrace\overline{u}^{(0)},\, \overline{v}^{(0)},\, 0,\, \overline{\tau}^{(0)},\, \overline{p}^{(0)}\right\}$ and three-dimensional $\left\{\overline{u},\, \overline{v},\, \overline{w},\, \overline{\tau},\, \overline{p}\right\rbrace$ gust signatures. The evolution of the former was considered by \citet{Ricco_2009} for the incompressible case, and is dominant in the outer part of the boundary layer. We focus on the three-dimensional velocity components because they dominate over the two-dimensional components as they exhibit the disturbance growth in the core of the boundary layer. Expanding the solution in terms of the three-dimensional gust signatures yields
\begin{equation}\label{3D_gust_signatures}
    \left\lbrace\overline{u}_0,\,\overline{v}_0,\,\overline{w}_0,\,\overline{\tau}_0,\,\overline{p}_0\right\rbrace 
    = 
    \left(\widehat{w}^\infty + \frac{\mathrm{i} k_z\widehat{v}^\infty}{\gamma}\right) 
    \left\lbrace\frac{\mathrm{i} k_z}{k_x}\overline{u},\, \mathrm{i} k_z\sqrt{\frac{2\overline{x}}{k_x\mathrm{R}_{\lambda}}}\overline{v},\,\overline{w},\,\frac{\mathrm{i} k_z}{k_x}\overline{\tau},\, \mathrm{i}\kappa_z \sqrt{\frac{k_x}{\mathrm{R}_\lambda}}\overline{p} \right\rbrace,
\end{equation}
where $\gamma=\left(k_x^2+k_z^2\right)^{1/2}$. Their evolution is governed by the compressible linearized unsteady boundary-region (CLUBR) equations \citepalias{Ricco_Wu_2007}. 

The CLUBR equations describe the evolution of the disturbances in the region III of \citetalias{Ricco_Wu_2007}, which occupies locations where $\eta=\mathcal{O}(1)$ and $\overline{x}=\mathcal{O}(1)$ downstream of the leading edge. The CLUBR equations are the limiting form of the compressible Navier-Stokes equations where the streamwise diffusion and the streamwise pressure gradient have been neglected. The boundary-layer thickness is comparable to $\lambda_z^\ast$ and the contribution of the spanwise diffusion to the momentum and energy balances must be taken into account. The wall-normal and spanwise diffusions are quantified by the asymptotic parameters

\begin{subequations}
\label{kappas}
\begin{align}
    \kappa_y&=\frac{k_y}{\sqrt{k_x\mathrm{R}_\lambda}} = \frac{2\pi}{\lambda_y^\ast}\left(\frac{\nu_\infty^\ast}{\omega^\ast}\right)^{1/2} =\mathcal{O}(1), \\
    \kappa_z&=\frac{k_z}{\sqrt{k_x\mathrm{R}_\lambda}} = \frac{2\pi}{\lambda_z^\ast}\left(\frac{\nu_\infty^\ast}{\omega^\ast}\right)^{1/2} =\mathcal{O}(1).
\end{align}
\end{subequations}
Free-stream gusts with equivalent wavenumbers $\kappa_y=\kappa_z$ are considered. The initial and boundary conditions are discussed in \S\ref{subsec:free-stream} and the modelling of the porous layer is presented in \S\ref{subsec:pores} and \S\ref{sec:admittance}. The CLUBR equations are given in Appendix \ref{app:CLUBR} \citep{Viaro_Ricco_2019a} and the details of their derivation are found in \cite{Ricco_2006}.

\subsubsection{Boundary and initial conditions}
\label{subsec:free-stream}

The CLUBR equations are subject to wall and free-stream boundary conditions that synthesize how the boundary layer interacts with the porous wall and the external disturbance flow. Being parabolic along $\overline{x}$, the CLUBR equations also require initial conditions for $\overline{x}\ll1$.

The no-slip wall boundary condition is applied to the streamwise and spanwise disturbance velocities, i.e. $\overline{u}=\overline{w}=0$ at $\eta=0$. At the wall, the wall-normal velocity and the temperature are related to the pressure because of the wall porosity, as follows

\begin{subequations}
\begin{align}
\overline{v}\left(\eta=0\right) &= \frac{\overline{A}_v}{(2 \overline{x})^{1/2}} \overline{p}\left(\eta=0\right),
\label{eq:bc-vel} \\
\overline{\tau}\left(\eta=0\right) &= \frac{\overline{A}_\tau}{(2 \overline{x})^{1/2}} \overline{p}\left(\eta=0\right), \label{eq:bc-temp}
\end{align}
\end{subequations}
where $\overline{A}_v$ and $\overline{A}_\tau$ are the scaled admittances, obtained in \S\ref{sec:admittance}.
The free-stream boundary conditions are the same as in \citetalias{Ricco_Wu_2007},

\begin{subequations}
\label{eq:br_bc_u_tau_lin}
\begin{align}
\left \lbrace \overline{u},\, \overline{\tau} \right \rbrace & \rightarrow 0 ,\\
\left(\frac{\p}{\p\eta} + \abs{\kappa_z} \left(2\overline{x}\right)^{1/2}\right) 
\left\lbrace \overline{v},\, \overline{w},\, \overline{p}\right\rbrace &
\rightarrow \left\lbrace -1,\, \mathrm{i} \kappa_y (2\overline{x})^{1/2},\, 0\right\rbrace e^{\mathrm{i}\left(\overline{x} + \kappa_y \left(2\overline{x}\right)^{1/2} \overline{\eta}\right)} e^{-\left(\kappa_y^2 + \kappa_z^2\right)\overline{x}},
\end{align}
\end{subequations}
as $\eta \rightarrow \infty$, where $\overline{\eta} \equiv \eta - \beta_c$, and $\beta_c = \lim_{\eta \rightarrow \infty} (\eta - F)$. The wall-normal wavenumber $\kappa_y$ only appears in \eqref{eq:br_bc_u_tau_lin} and not in the CLUBR equations because the wall-normal length scale of the free-stream flow is $\lambda^\ast_y$, while, within the boundary layer, the characteristic length scale is the boundary-layer thickness. 

The initial conditions are the same as in \citetalias{Ricco_Wu_2007}. As they pertain to a non-porous wall, the wall porosity increases smoothly from zero at small $\overline{x}$ to a finite value downstream according to the function proposed by \cite{Negi_Mishra_Skote_2015}, as discussed in \S\ref{sec:admittance}. A few comments about the initial conditions are in order. The plate is assumed to be infinitely thin and therefore the free-stream base flow is not distorted at leading order as the fluid encounters the flat plate. The only distortion of the base-flow streamlines is produced by the thickening of the boundary layer. As the free-stream disturbances are transported by the base flow, they are neither stretched nor tilted by the leading edge. 
The leading-edge bluntness effects can play a central role on the free-stream distortion and therefore on the boundary-layer response. This problem is however out of the scope of the present study because these effects only occur when the characteristic dimension of the rounded leading edge is comparable with the spanwise length scale \citep{goldstein-leib-cowley-1992,goldstein-leib-1993,goldstein-wundrow-1998,goldstein-2014}. 
Furthermore, the disturbance flow in the very proximity of the leading edge is not considered because the inviscid flow outside of the boundary layer is solved for $x \gg 1$, i.e. at a distance much larger than the spanwise wavelength. As discussed in LWG99, streamwise-decaying disturbances emerging from the interaction between the free-stream vorticity disturbances and the leading edge, obtained by \cite{Choudhari_1996} by using the Wiener-Hopf technique, decay to a very small amplitude when $x \gg 1$ and, therefore, they play a negligible role in the boundary-layer response. As the initial conditions are obtained by taking the limit $\overline{x} \ll 1$ of the CLUBR equations, they constitute the asymptotically rigorous upstream behaviour of the CLUBR solution at locations $1 \ll x \ll k_x^{-1}$ or, in dimensional form, at locations $\lambda_z^* \ll x^* \ll \lambda_x^*$.

The base-flow solutions \eqref{base_flow_solutions} are computed using a second-order accurate Keller-box method \citep{Cebeci_2002}. The CLUBR system, given in Appendix \ref{app:CLUBR}, is solved by a second-order finite-difference scheme that is central in $\eta$ and backward in $\overline{x}$. A standard block-elimination algorithm is utilized \citep{Cebeci_2002}. The free-stream boundary conditions \eqref{eq:br_bc_u_tau_lin} are applied by a second-order finite-difference discretization scheme. The pressure is computed on a grid staggered along the $\eta$ direction with respect to that for the velocity in order to avoid the pressure decoupling phenomenon. 

\subsubsection{The unsteady disturbance flow within the pores}
\label{subsec:pores}
The flow inside a pore in studied in this section. In the porous wall designed by \cite{Fedorov_Malmuth_Rasheed_Hornung_2001}, the pressure fluctuations at the interface between the wall and the boundary layer excite kinematic and thermal disturbances in long, thin cylindrical pores. The numerical studies of \citet{Zhao_Liu_Wen_Zhu_Cheng_2018,Zhao_Zhang_Wen_2020} showed that the effects of acoustic scattering between adjacent pores can be neglected when the Helmholtz number $\mathrm{He}=\omega^\ast H^\ast/c_w^\ast<4.21$, where $H^\ast$ is the depth of the pores and $c_w^\ast=\sqrt{\gamma\mathcal{R}^\ast T_w^\ast}$ the speed of sound in the pores. All the cases considered in the present work comply with that condition. Hence, the properties of the porous layer can be studied by considering the flow characteristics of an isolated pore. The equations that govern the propagation of small-amplitude disturbances in a single dead-end circular pore of depth $H^\ast$ and radius $R^\ast$ are reported in Appendix \ref{app:appendix_admittance}. The linearized continuity, axial momentum and energy  equations are cast in cylindrical coordinates, and their solution yields an analytical radial distribution of the velocity and temperature in the form of Bessel functions \citep{Zwikker_Kosten_1949,Fedorov_Malmuth_Rasheed_Hornung_2001}. 

\begin{figure}
    \centering
    \begin{subfigure}[b]{0.49\textwidth}
        \includegraphics[width=\textwidth]{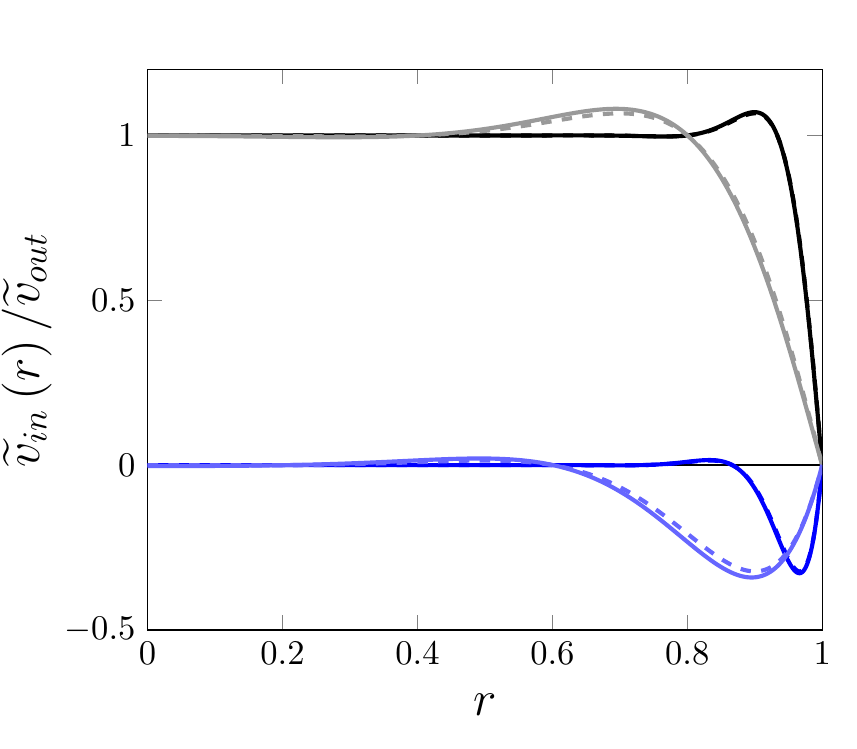}
        \put(-200,180){a)}
    \end{subfigure}
    \begin{subfigure}[b]{0.49\textwidth}
        \includegraphics[width=\textwidth]{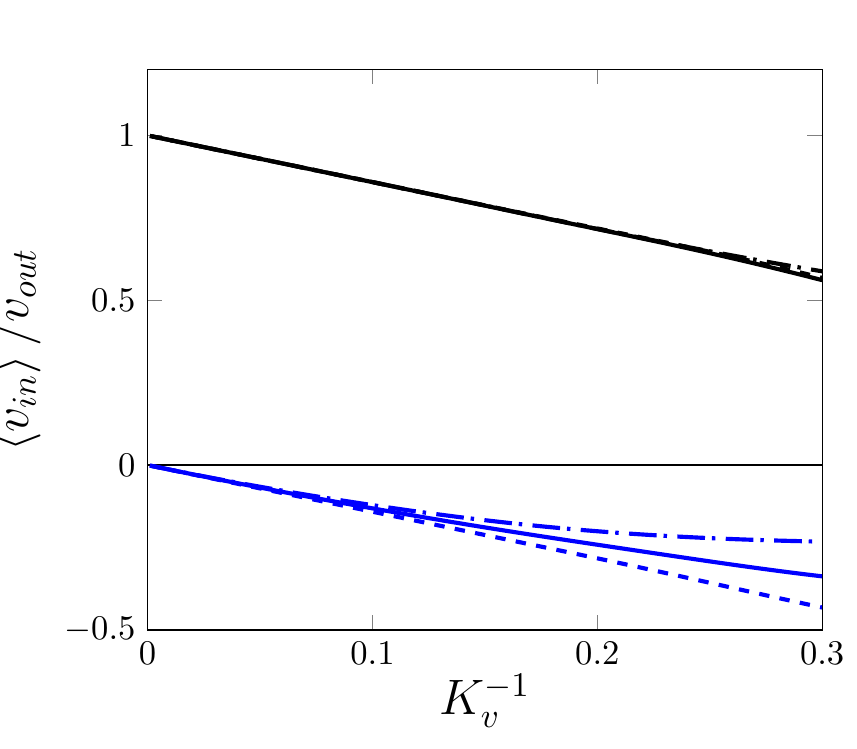}
        \put(-200,180){b)}
    \end{subfigure}
    \caption{(a) Radial distribution of the axial velocity within a pore at a given depth for $K_v=11$ (grey for real part and light blue for imaginary part) and $K_v=33$ (black for real part and dark blue for imaginary part). The Bessel-function solutions (\ref{REG_POROUS_cylindrical_solutions_velocity}, solid lines) are compared with the asymptotic solutions (\ref{inner_sol_vel}, dashed lines). (b) Averaged Bessel-function solutions (\ref{REG_POROUS_cylindrical_solutions_velocity}, solid lines), averaged asymptotic solutions (\ref{inner_sol_vel}, dashed curves), and averaged Bessel-function solutions obtained with the asymptotic form of the Bessel functions for large arguments (\ref{bessel_large_arguments}, dash-dotted lines) for $K_v=11$ (blue lines) and $K_v=33$ (black lines).}
    \label{fig:tube_asymptotics}
\end{figure}

The response of the pores is ruled by the frequency parameter $k_x\mathrm{R}_\lambda$ \citep{Fedorov_Malmuth_Rasheed_Hornung_2001,Goldstein_Ricco_2018,Viaro_Ricco_2019a}. The disturbances are not transmitted to the pores at very low frequencies, for which the frequency parameter $k_x\mathrm{R}_\lambda =\Or{1}$ or smaller. The boundary-layer disturbances are expected to interact with the porous wall as $k_x\mathrm{R}_\lambda$ increases and the magnitude of the spanwise diffusion, proportional to $\kappa_z$ in \eqref{kappas}, decreases. We are therefore interested in investigating the behaviour of the porous layer for $k_x\mathrm{R}_\lambda \gg 1$. The terms of the momentum and energy balances in \eqref{REG_POROUS_cylindrical} are scaled as in the boundary layer and the parameter

\begin{equation}
\label{Kv_parameter}
    K_v = R^\ast\left(\frac{\rho_w^\ast\omega^\ast}{\mu_w^\ast}\right)^{1/2} = R\left(\frac{k_x\mathrm{R}_\lambda}{\mu_wT_w}\right)^{1/2}\gg 1
\end{equation}
is introduced, where $\rho_w^\ast$ and $\mu_w^\ast$ are the base-flow density and dynamic viscosity at the wall. The balance equations reveal a boundary-layer structure \citep{Bender_Orszag_1999} for the velocity and temperature fluctuations, whose values depend on the radial coordinate $r$, while the pressure is only a function of the axial coordinate $y$ and is the same inside and outside the boundary layer. The outer solutions are obtained by imposing $K_v^{-1}=0$, for which the full system \eqref{REG_POROUS_cylindrical} in Appendix \ref{app:appendix_admittance} reduces to the equation for the pressure

\begin{equation}
\label{pseudo-helmholtz-equation}
    \frac{\mathrm{d}^2\widetilde{p}}{\mathrm{d} y^2} + \mathrm{He}^2\widetilde{p} =0,
\end{equation}
which arises from a reduction of a Helmholtz equation. The outer solutions for the velocity and pressure fluctuations are found,

\begin{subequations}
\begin{align}
    \widetilde{p} = \widetilde{p}_{out}\left(y\right) &= a\cos{\left[\mathrm{He}\left(y+1\right)\right]}, \\
    \widetilde{v} = \widetilde{v}_{out}\left(y\right) &= -\mathrm{i}\frac{\mathrm{d}\widetilde{p}}{\mathrm{d} y} = \mathrm{i} a\sin{\left[\mathrm{He}\left(y+1\right)\right]},
    \label{REG_POR_outer_velocity}\\
    \widetilde{\tau} = k_x^2 \widetilde{\tau}_{out}\left(y\right) &= k_x^2\left(\gamma-1\right)a\frac{\mathrm{M}_\infty^2H^2}{T_w}\cos{\left[\mathrm{He}\left(y+1\right)\right]},
    \label{REG_POR_outer_temperature} 
\end{align}
\end{subequations}
where $a$ is a real constant. The pressure and temperature fluctuations are in phase in the outer region. In the proximity of the wall, i.e. where $r-1 \ll 1$, an inner variable 

\begin{equation}
    r_s = K_v\left(1-r\right) = \mathcal{O}(1)
\end{equation}
describes the inner solutions $\widetilde{v}_{in}\left(r_s,y\right)$ and $k_x^2\widetilde{\tau}_{in}\left(r_s,y\right)$. Upon introduction of the inner variable, the momentum and energy balance equations take the form

\begin{subequations}
\label{inner_eq}
\begin{align}
    -\mathrm{i}\widetilde{v}_{in} + \frac{\mathrm{d}\widetilde{p}_{out}}{\mathrm{d}y} &= \frac{\p^2\widetilde{v}_{in}}{\p r_s^2}, \\
    -\mathrm{i}\widetilde{\tau}_{in} + \mathrm{i}\left(\gamma-1\right)\frac{\mathrm{M}_\infty^2L^2}{T_w}\widetilde{p}_{out} &= \frac{1}{\mathrm{Pr}}\frac{\p^2\widetilde{\tau}_{in}}{\p r_s^2},
\end{align}
\end{subequations}
subject to the boundary conditions

\begin{subequations}
\begin{align}
    \widetilde{v}_{in}\left(0;y\right) = \widetilde{\tau}_{in}\left(0;y\right) &= 0,\\
    \lim_{r_s\to\infty}\widetilde{v}_{in}\left(r_s;y\right) &= \widetilde{v}_{out}\left(y\right), \\
    \lim_{r_s\to\infty}\widetilde{\tau}_{in}\left(r_s;y\right) &= \widetilde{\tau}_{out}\left(y\right).
\end{align}
\end{subequations}
The inner solutions are

\begin{subequations}
\label{inner_sol}
\begin{align}
    \widetilde{v}_{in}\left(r;y\right) &= \widetilde{v}_{out}\left(y\right)\left[1-\exp\left(\mathrm{i}^{3/2}K_v\left(r-1\right)\right)\right], 
    \label{inner_sol_vel}\\
    \widetilde{\tau}_{in}\left(r;y\right) &= \widetilde{\tau}_{out}\left(y\right)\left[1-\exp\left(\mathrm{i}^{3/2}\mathrm{Pr}^{1/2}K_v\left(r-1\right)\right)\right].
    \label{inner_sol_temp}
\end{align}
\end{subequations}
The solutions \eqref{inner_sol} represent azimuthal Stokes layers of velocity and temperature attached to the pore wall. The cross-sectional averages of the axial velocity \eqref{inner_sol_vel} and the temperature perturbations \eqref{inner_sol_temp} over the circular section of a pore are

\begin{subequations}
\label{inner_sol_avg}
\begin{align}
    \left\langle\widetilde{v}_{in}\right\rangle\left(r;y\right) &= \widetilde{v}_{out}\left(y\right)\left[1+2\mathrm{i}\frac{\displaystyle \exp\left(-\mathrm{i}^{3/2}K_v\right) + \mathrm{i}^{3/2}K_v - 1}{K_v^2}\right], \\
    \left\langle\widetilde{\tau}_{in}\right\rangle\left(r;y\right) &= \widetilde{\tau}_{out}\left(y\right)\left[1 + 2\mathrm{i}\frac{\displaystyle \exp\left(-\mathrm{i}^{3/2}\mathrm{Pr}^{1/2}K_v\right) + \mathrm{i}^{3/2}\mathrm{Pr}^{1/2}K_v-1}{\mathrm{Pr}K_v^2}\right] .
\end{align}
\end{subequations}
Figure \ref{fig:tube_asymptotics}a shows that the agreement between the Bessel-function solutions, given in \eqref{REG_POROUS_cylindrical_solutions} of Appendix \ref{app:appendix_admittance}, and the asymptotic solution \eqref{inner_sol_vel} improves as $K_v$ increases, the lines being indistinguishable for $K_v=33$. For $K_v \gg 1$, the cross-sectional average of the Bessel-function axial velocity \eqref{REG_POROUS_cylindrical_averages_velocity} can also be obtained by using the asymptotic expansion for large arguments of the Bessel function. The leading order terms \citep{Abramowitz_Stegun_1970}

\begin{equation}
\label{bessel_large_arguments}
    J_m\left(\xi\right) = \left(\frac{2}{\pi \xi}\right)^{1/2}\cos{\left(\xi-\frac{m\pi}{2}-\frac{\pi}{4}\right)}   \mbox{, }m=0,1,2,\dots
\end{equation}
yield the relation

\begin{equation}\label{REG_POROUS_cylindrical_averages_velocity_large_arg}
    \average{\widetilde{v}_{in}}\left(y\right) = \widetilde{v}_{out}\left[1-\frac{\cos{\left(\mathrm{i}^{1/2}K_v-3\pi/4\right)}}{\mathrm{i}^{1/2}K_v\cos{\left(\mathrm{i}^{1/2}K_v-\pi/4\right)}}\right].
\end{equation}

The real and imaginary parts of \eqref{REG_POROUS_cylindrical_averages_velocity}, \eqref{REG_POROUS_cylindrical_averages_velocity_large_arg}, and \eqref{inner_sol} are normalized with respect to $\widetilde{v}_{out}$ and plotted in figure \ref{fig:tube_asymptotics}b. The difference between the real parts becomes indiscernible for $K_v > 4$, whereas the imaginary parts match excellently for $K_v>10$. The ratio 

\begin{equation}\label{normalized_admittance_definition}
    \frac{\average{\widetilde{v}_{in}}}{\widetilde{v}_{out}} = A_v\frac{\widetilde{p}}{\widetilde{v}_{out}}
\end{equation}
represents a normalized acoustic admittance, where the normalization factor is the large-$K_v$ limit of $A_v$.

\subsubsection{Porous admittance}
\label{sec:admittance}

\begin{figure}
    \centering
    \begin{subfigure}[b]{0.49\textwidth}
        \includegraphics[width=\linewidth]{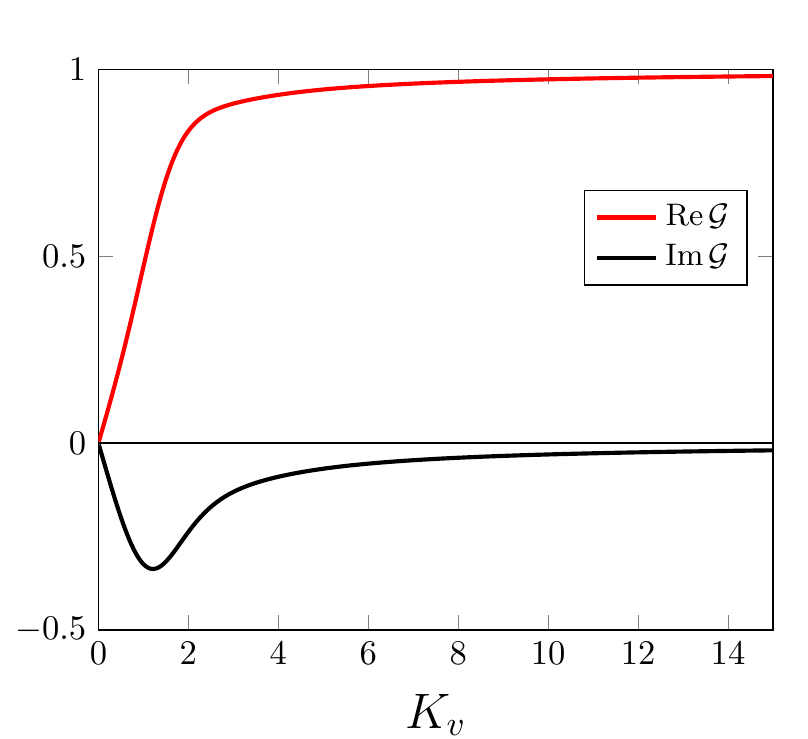}
        \put(-200,180){a)}
    \end{subfigure}
    \begin{subfigure}[b]{0.49\textwidth}
        \includegraphics[width=\linewidth]{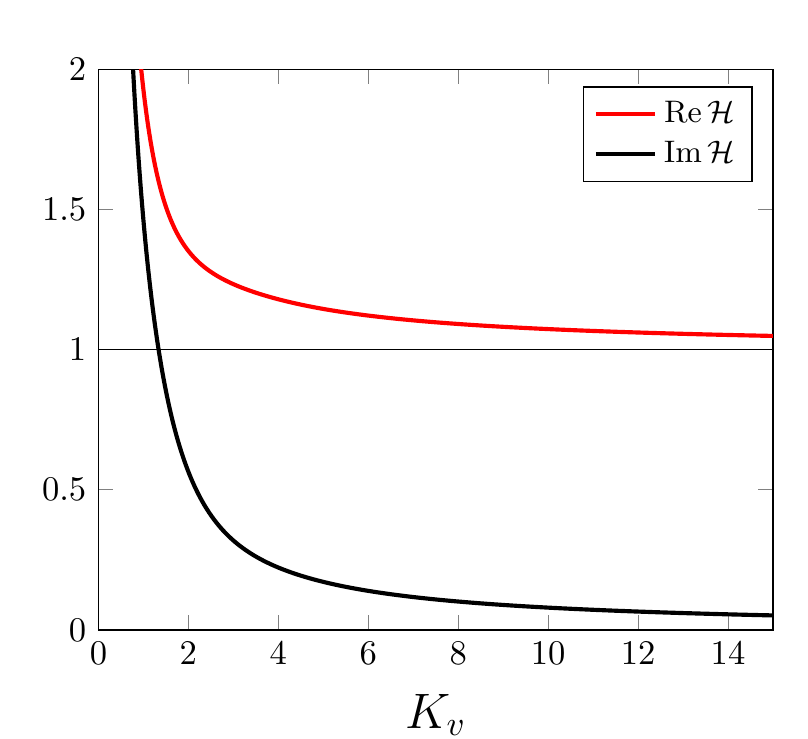}
        \put(-200,180){b)}
    \end{subfigure}
    \caption{Real and imaginary parts of $\mathcal{G}$ and $\mathcal{H}$ as functions of $K_v$, given by the Bessel-function solutions \eqref{G_function} and \eqref{H_function}, respectively.}
    \label{fig:complex_functions}
\end{figure}

Following \cite{Fedorov_Malmuth_Rasheed_Hornung_2001}, the wall-normal velocity disturbance and the temperature disturbance are related to the pressure disturbance as follows,

\begin{subequations}
\begin{align}
    \widetilde{v}\left(\eta=0\right) &= A_v \widetilde{p}\left(\eta=0\right), \\
    \widetilde{\tau}\left(\eta=0\right) &= A_\tau \widetilde{p}\left(\eta=0\right) ,
\end{align}
\end{subequations}
where $A_v$ and $A_\tau$ are the complex admittances of the porous wall evaluated at the wall-boundary layer interface ($\eta=0$). They are derived in Appendix \ref{app:appendix_admittance}. The velocity admittance is

\begin{equation}\label{admittance_velocity_early}
    A_v=-\phi\frac{\mathrm{i}\Lambda}{L}\left[1-\mathcal{F}\left(\mathrm{i}^{1/2}K_v\right)\right]\tanh{\Lambda},
\end{equation}
where

\begin{equation}\label{propagation_constant}
    \Lambda = \frac{\mathrm{i} k_x\mathrm{M}_\infty L}{T_w^{1/2}}\mathcal{H}\left(\mathrm{i}^{1/2}K_v\right),
\end{equation}

\begin{equation}
\label{H_function}
    \mathcal{H}\left(\mathrm{i}^{1/2}K_v\right) = \left[\frac{1+\left(\gamma-1\right)\mathcal{F}\left(\left(\mathrm{i}\mathrm{Pr}\right)^{1/2}K_v\right)}{1-\mathcal{F}\left(\mathrm{i}^{1/2}K_v\right)}\right]^{1/2},
\end{equation}
and $\mathcal{F}$ is given by \eqref{eq:F-hat}. The porosity $\phi$ is defined as the ratio between the surface area of the pores and the total surface area. By combining \eqref{admittance_velocity_early} and \eqref{propagation_constant}, the admittance of the velocity is rewritten as 

\begin{equation}
\label{velocity_admittance}
    A_v = \phi\frac{k_x\mathrm{M}_\infty}{T_w^{1/2}}\mathcal{G}\left(\mathrm{i}^{1/2}K_v\right)\tanh{\Lambda},
\end{equation}
where

\begin{equation}
\label{G_function}
    \mathcal{G}\left(\mathrm{i}^{1/2}K_v\right) = \left[\left(1-\mathcal{F}\left(\mathrm{i}^{1/2}K_v\right)\right)\left(1+\left(\gamma-1\right)\mathcal{F}\left(\left(\mathrm{i}\mathrm{Pr}\right)^{1/2}K_v\right)\right)\right]^{1/2}.
\end{equation}
Figures \ref{fig:complex_functions}a and \ref{fig:complex_functions}b show the real and imaginary parts of $\mathcal{G}$ and $\mathcal{H}$, respectively.

The thermal admittance \eqref{REG_POROUS_admittance_temperature} reads
\begin{equation}
\label{eq:thermal_admittance}
    A_\tau=\left(\gamma-1\right)\frac{k_x^2\mathrm{M}_\infty^2L^2}{T_w}\left[1-\mathcal{F}\left(\left(\mathrm{i}\mathrm{Pr}\right)^{1/2}K_v\right)\right].
\end{equation}
By introducing the expressions of the boundary-layer disturbances \eqref{3D_gust_signatures} and by using \eqref{kappas}, one finds

\begin{subequations}
\begin{align}
\overline{v}\left(\eta=0\right) &= \frac{k_x \kappa_z A_v}{(2 \overline{x})^{1/2} k_z} \overline{p}\left(\eta=0\right) = \left(\frac{k_x}{2\overline{x}\mathrm{R}_\lambda}\right)^{1/2} A_v \overline{p}\left(\eta=0\right),
\label{eq:bc_velocity} \\
\overline{\tau}\left(\eta=0\right) &= \frac{k_x A_\tau}{\mathrm{R}_\lambda} \overline{p}\left(\eta=0\right). \label{eq:bc_temperature}
\end{align}
\end{subequations}

\sloppypar{
The wall boundary conditions \eqref{eq:bc_velocity} and \eqref{eq:bc_temperature} are rewritten by using $\overline{A}_v=\left(k_x\mathrm{R}_\lambda^{-1}\right)^{1/2}A_v$ and $\overline{A}_\tau=\left(k_x\mathrm{R}_\lambda^{-1}\right)A_\tau$ in \eqref{eq:bc-vel} and \eqref{eq:bc-temp}.
The velocity admittance $A_v$ defined in \eqref{velocity_admittance} is $\Or{k_x}$ and the thermal admittance $A_\tau$ \eqref{eq:thermal_admittance} is $\Or{k_x^2}$. The coefficients in front of the pressure in \eqref{eq:bc_velocity} and \eqref{eq:bc_temperature} are $\mathcal{O}\left(k_x^{3/2}\mathrm{R}_\lambda^{-1/2}\right)$ and $\mathcal{O}\left(k_x^3\mathrm{R}_\lambda^{-1}\right)$, respectively. The contribution of $A_\tau$ to the temperature fluctuations is thus much weaker than the contribution of $A_v$ to the velocity fluctuations and therefore negligible. However, the porous wall affects the temperature fluctuations indirectly because of the coupling between the wall-normal momentum equation and the energy equation. For typical Klebanoff modes and G\"{o}rtler vortices $k_x\mathrm{R}_\lambda=\order{1}$, $A_v=\order{\mathrm{R}_\lambda^{-2}}$, $A_v=\order{\mathrm{R}_\lambda^{-4}}$, and wall porosity has a negligible effect on both the velocity and temperature fluctuations. As shown in \S\ref{subsec:pores} the pores begin interacting with the disturbance flow when $k_x\mathrm{R}_\lambda\gg1$. Since the pressure and temperature fluctuations are in phase within the pores, the adiabatic boundary condition can be imposed at the wall in accordance with the homogeneous Neumann boundary condition at the dead end of the pores, as discussed in Appendix \ref{app:appendix_admittance}.

Since the upstream boundary conditions for $\overline{x}\ll 1$ are not compatible with a non-zero wall-normal velocity at $\eta =0$, a short smoothing region along the streamwise direction is introduced between two streamwise coordinates $\overline{x}_1$ and $\overline{x}_2$ in the vicinity of the leading edge. In this region, the velocity admittance varies proportionally to \citep{Negi_Mishra_Skote_2015}
\begin{equation}
\label{skote_function}
  S (\overline{x}) =
    \begin{cases}
      0,   & \text{for \ $\overline{x} \leq \overline{x}_1$},\\
      \left[ 1 + \exp\left(\displaystyle{\frac{1}{ \widetilde x - 1 } + \frac{1}{ \widetilde x}}\right) \right]^{-1}, & \text{for \ $\overline{x}_1 < \overline{x} < \overline{x}_2$},\\
      1, & \text{for $\overline{x} \geq \overline{x}_2$},
    \end{cases}       
\end{equation}
where $\widetilde x = (\overline{x} - \overline{x}_1)/(\overline{x}_2 - \overline{x}_1)$, and $\overline{x}_1=0.005$. The end point is $\overline{x}_2=0.01$, if exception is made for the analysis at the end of \S\ref{sec:k-modes}, where the effect of $\overline{x}_2$ is studied. The piecewise function \eqref{skote_function} can be physically interpreted as a variation of the wall porosity along the smoothing region. If we assume the pores to be aligned in equally spaced rows and columns, such variation may be caused by pores of constant radius $R^\ast$ becoming more and more packed as the distance $d^\ast$ between the centres of adjacent pores decreases, or by the gradual increase of the pore radius. In both cases, the porosity in \eqref{velocity_admittance} can be written as $\phi = \pi R_f^{\ast2} S(\overline{x})/d_f^{\ast2}$, where the subscript $f$ denotes quantities at the downstream end of the smoothing region. If the radius is kept constant between $\overline{x}_1$ and $\overline{x}_2$, the interpore distance is $d^\ast(\overline{x}) = d_f^\ast/\sqrt{S(\overline{x})}$. The porosity at the end of the smoothing region is thus $\phi=\pi R_f^{\ast2}/d_f^{\ast2}$. As only regularly-spaced circular pores are considered, the porosity may attain a maximum theoretical value of $\pi/4$ when $R_f^\ast=d_f^\ast/2$. 

\section{Results}
\label{sec:results}
The effectiveness of the porous wall depends on its ability to transduce a pressure disturbance into a wall-normal velocity disturbance, as described by \eqref{eq:bc_velocity}. In the present coatings, the phase velocity of the disturbances is equivalent to the local sound speed \citep{Bres_Inkman_Colonius_Fedorov_2013}. For $K_v\gg 1$, a dimensional analysis of the boundary condition \eqref{eq:bc_velocity} and the velocity admittance \eqref{velocity_admittance} yields
\begin{equation}
    \frac{\overline{v}}{\overline{p}} = \Or{\left(\frac{k_x^3}{R_\lambda}\right)^{1/2}\frac{\mathrm{M}_\infty}{T_w^{1/2}}} = \Or{\frac{\omega^{\ast3/2}\lambda_z^\ast}{U_\infty^\ast c_w^\ast}\nu_\infty^{\ast 1/2}}.
\end{equation}
As a result, the pores interact with the boundary layer when
\begin{equation}
\label{porous_layer_effectiveness}
    \omega^{\ast3/2}\lambda_z^\ast \mbox{ is comparable with } U_\infty^\ast c_w^\ast\nu_\infty^{\ast-1/2}.
\end{equation}
A visual representation of relation \eqref{porous_layer_effectiveness} is shown in figure \ref{fig:porosity-regimes}. The free-stream velocity, the free-stream kinematic viscosity and the speed of sound in the porous layer define the threshold above which the boundary-layer disturbances are affected by the porous layer. For given free-stream conditions, the effect of the porous layer is more intense at lower $c_w^\ast$. For a given $\lambda_z^\ast$ and constant free-stream conditions, the minimum frequency at which a disturbance is affected grows as $T_w^{\ast1/3}$. The minimum wavelength for which a disturbance is attenuated at a given frequency increases as $T_w^{\ast1/2}$. Both hyperbolas shift away from the origin as $c_w^\ast$ increases.
Relation \eqref{porous_layer_effectiveness} also shows that the variation of the frequency is more influential on the performance of the porous wall than that of the spanwise wavelength. 

\begin{figure}
    \centering
    \includegraphics[width=0.65\linewidth]{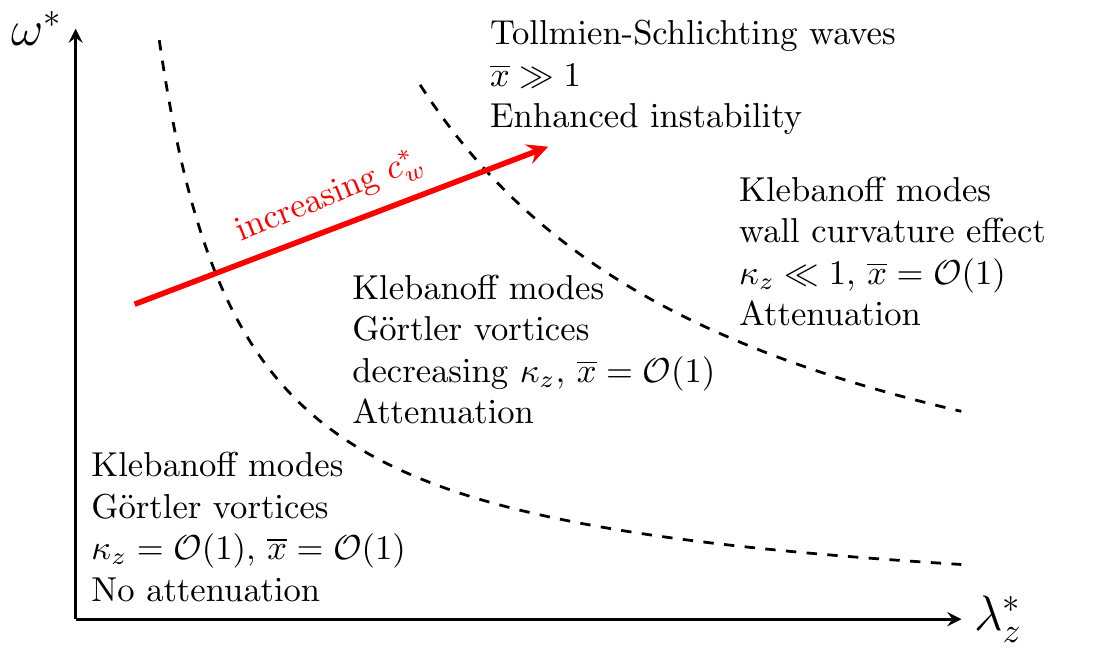}
    \caption{Schematic of flow regimes and the effect of wall porosity, as described by relation \eqref{porous_layer_effectiveness}.} 
    \label{fig:porosity-regimes}
\end{figure}

\begin{table}
\centering
\begin{tabular}{ |p{5.5cm} P{2.5cm} P{3cm} P{1.5cm}|  }
 \hline
 \textbf{Physical parameter} & \textbf{Symbol} & \textbf{Value} & \textbf{SI unit}\\
 Mach number & $\mathrm{M}_\infty$ & $6$ & - \\
 Total (stagnation) temperature & $T_0^\ast$ & 400 & $\si{\kelvin}$
\\
 Static pressure & $p_\infty^\ast$ & 633 & $\si{\pascal}$ \\
 Static temperature & $T_\infty^\ast$ & 49 & $\si{\kelvin}$
\\
 Free-stream velocity & $U_\infty^\ast$ & 841 & $\si{\meter\per\second}$ \\
 Free-stream kinematic viscosity & $\nu_\infty^\ast$ &  $6.3\cdot10^{-5}$ & $\si{\meter^2\per\second}$ \\
 Unit Reynolds number & $\mathrm{R}^\ast=U_\infty^\ast/\nu_\infty^\ast$ &  $13.5\cdot10^6$ & $\si{\per\meter}$\\
 Recovery temperature & $T^\ast_{a,w}$ & 343 & $\si{\kelvin}$ \\
 Wall temperature & $T_w^\ast=0.8T^\ast_{a,w}$ & 274 & $\si{\kelvin}$ \\
 Pore radius & $R^\ast$  & $90$ & $\si{\micro\meter}$ \\
 Inter-pore distance & $d^\ast$ & $210$ & $\si{\micro\meter}$ \\
 Pore depth & $H^\ast$ & $1.5$ & $\si{\milli\meter}$ \\
 Porosity & $\phi$ & $0.58$ & - \\
     Velocity admittance & $\overline{A}_v$ & $-8.82\cdot10^{-4}+1.434\cdot10^{-3} \ \mathrm{i}$ & - \\
\hline
\end{tabular}
\caption{Physical parameters for wind tunnel conditions. \label{tab:physical_quantities}} 
\end{table}

The physical parameters of the present study are listed in table \ref{tab:physical_quantities}. These values are representative of supersonic quiet tunnel conditions, such as those of the Sandia Hypersonic Wind tunnel and the Boeing Mach 6 quiet tunnel \citep{Casper_Beresh_Henfling_Spillers_Pruett_Schneider_2009}. The stagnation temperature of $400\,\si{\kelvin}$ and the wall-temperature ratio of $T_w^\ast/T_{ad,w}^\ast=0.8$ (where $T_{ad,w}^*$ is the adiabatic-wall temperature) are given by \cite{Shiplyuk_Burov_Maslov_Fomin_2004,Casper_Beresh_Henfling_Spillers_Pruett_Schneider_2009,Schneider_2008} and \cite{Yu_Xu_Liu_Zhang_2018}.

\subsection{Klebanoff modes}
\label{sec:k-modes}

\begin{figure}
\centering 
    \begin{subfigure}[b]{0.49\textwidth}
        \includegraphics[width=\linewidth]{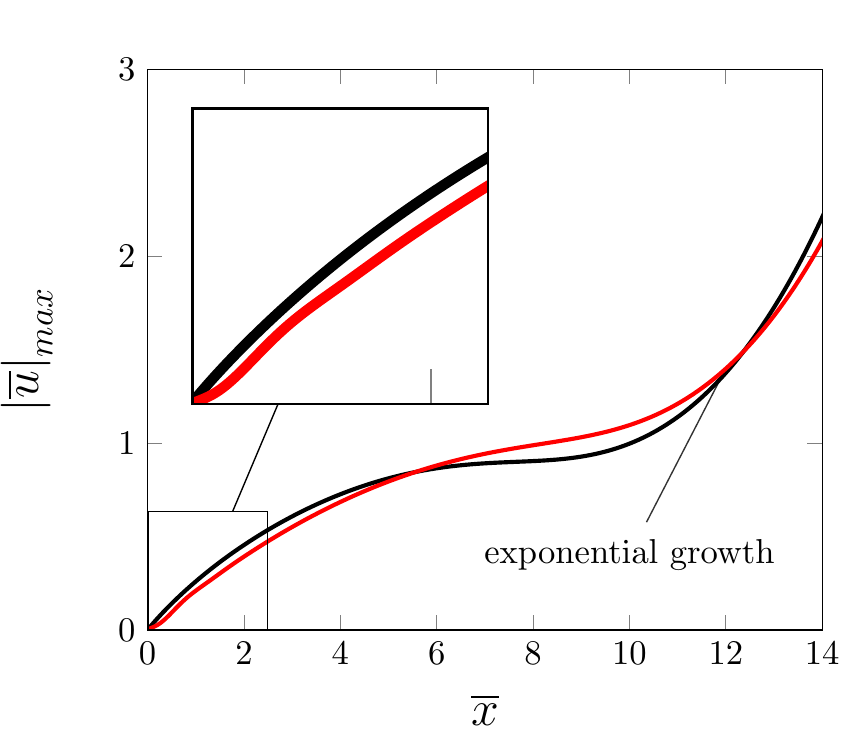}
        \caption{$\kappa_z=0.008$, $k_x/k_z=0.24$.}
        \label{fig:effect-frequency-1}
    \end{subfigure}
    \begin{subfigure}[b]{0.49\textwidth}
        \includegraphics[width=\linewidth]{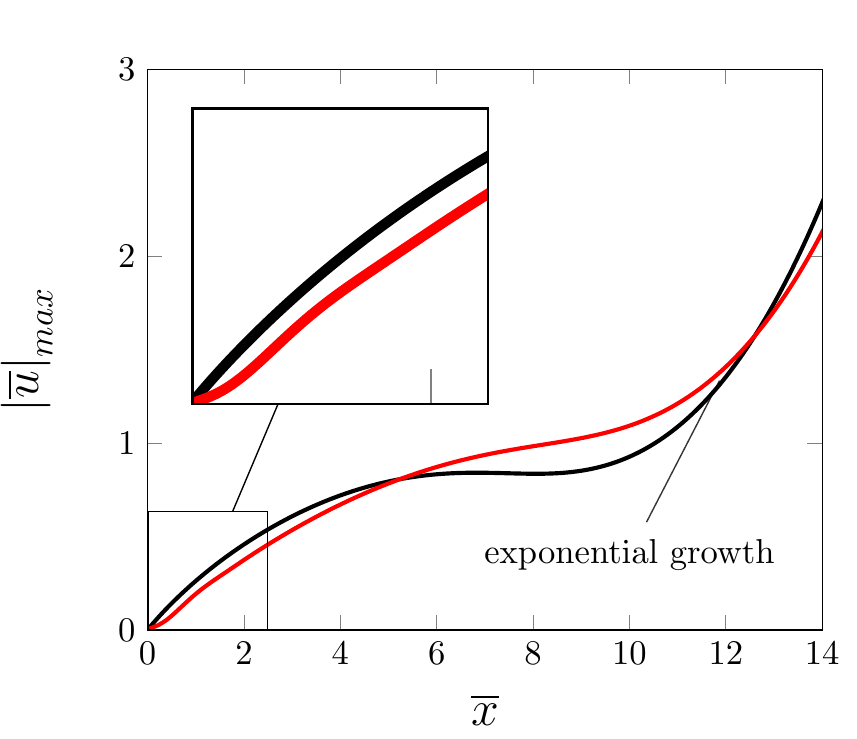}
        \caption{$\kappa_z=0.0075$, $k_x/k_z=0.27$.}
        \label{fig:effect-frequency-2}
    \end{subfigure}
    \begin{subfigure}[b]{0.49\textwidth}
        \includegraphics[width=\linewidth]{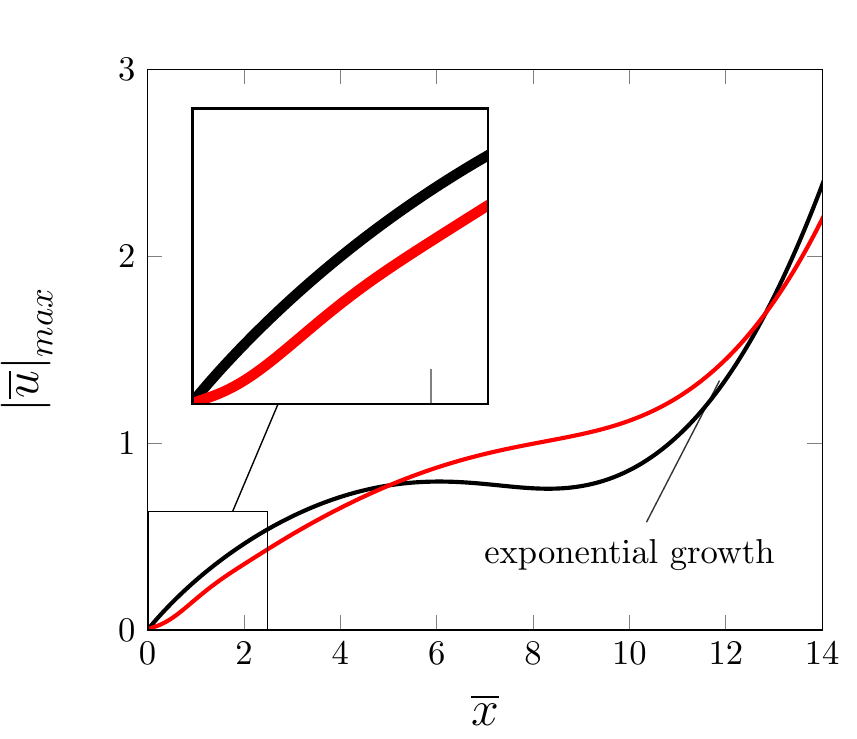}
        \caption{$\kappa_z=0.007$, $k_x/k_z=0.31$.}
        \label{fig:effect-frequency-3_umax}
    \end{subfigure}
    \begin{subfigure}[b]{0.49\textwidth}
        \includegraphics[width=\linewidth]{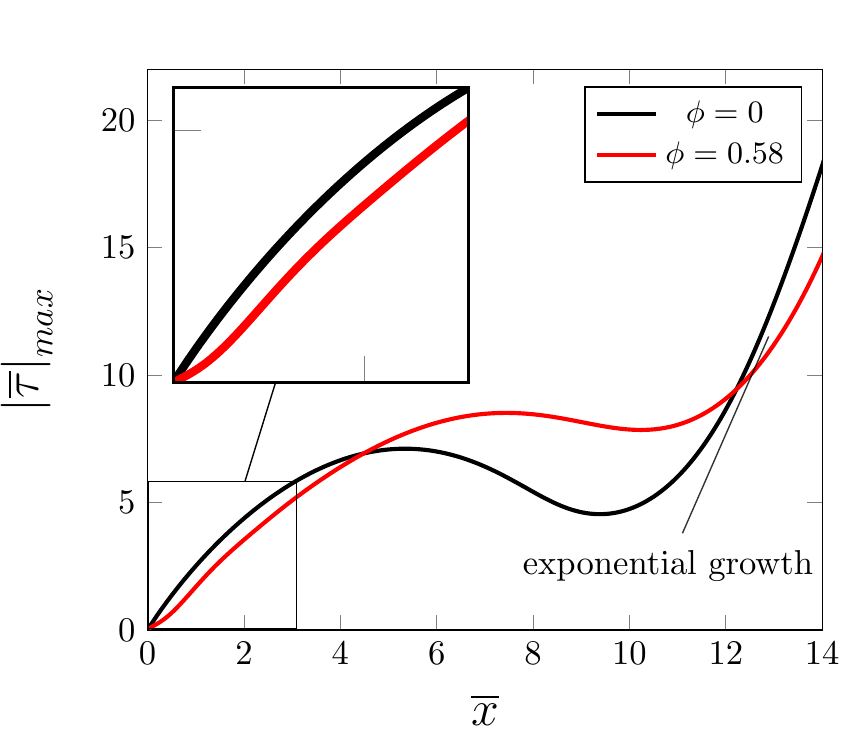}
        \caption{$\kappa_z=0.007$, $k_x/k_z=0.31$.}
        \label{fig:effect-frequency-3_taumax}
    \end{subfigure}
\caption{Effect of the frequency on the attenuation streamwise velocity (figures \ref{fig:effect-frequency-1}, \ref{fig:effect-frequency-2}, \ref{fig:effect-frequency-3_umax}) and temperature \ref{fig:effect-frequency-3_taumax}) for  $\lambda_z^\ast=0.03\,\si{\meter}$ ($\mathrm{R}_\lambda=418000$) at $\omega^\ast/2\pi = 6730\,\si{\hertz}$ (\ref{fig:effect-frequency-1}), $\omega^\ast/2\pi = 7570\,\si{\hertz}$ (\ref{fig:effect-frequency-2}) and $\omega^\ast/2\pi = 8600\,\si{\hertz}$ (\ref{fig:effect-frequency-3_umax}-\ref{fig:effect-frequency-3_taumax}). The solid ($\phi=0$) and porous ($\phi=0.58$) wall cases are represented with black and red curves, respectively.} 
\label{fig:effect-frequency}
\end{figure}

\begin{figure}
    \centering
    \begin{subfigure}[b]{0.49\textwidth}
        \includegraphics[width=\linewidth]{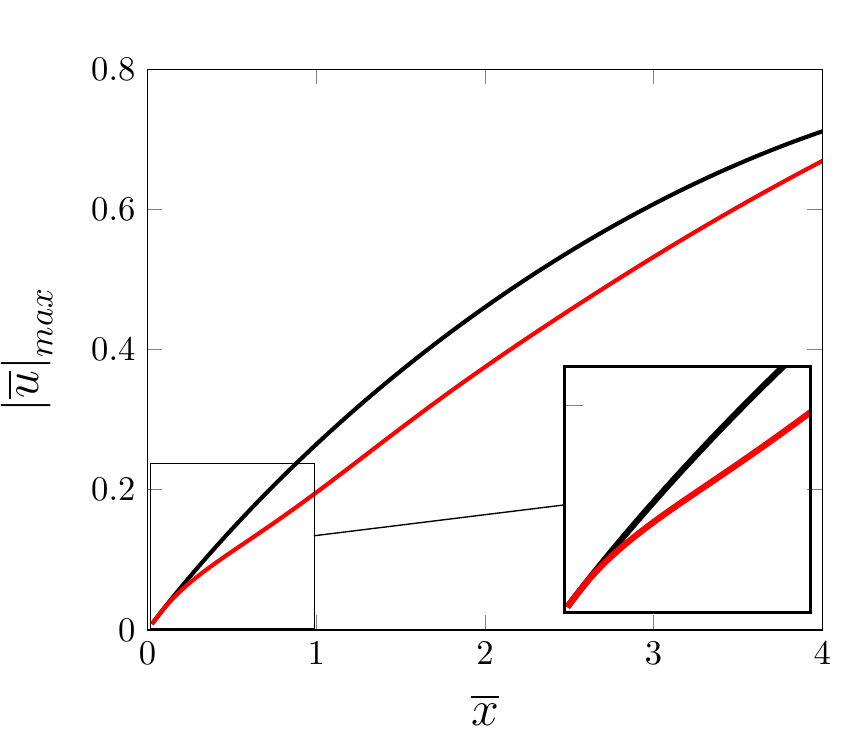}
        \caption{$\overline{x}_2=0.5$}
        \label{fig:effect_transition_layer_skote05}
    \end{subfigure}
    \begin{subfigure}[b]{0.49\textwidth}
        \includegraphics[width=\linewidth]{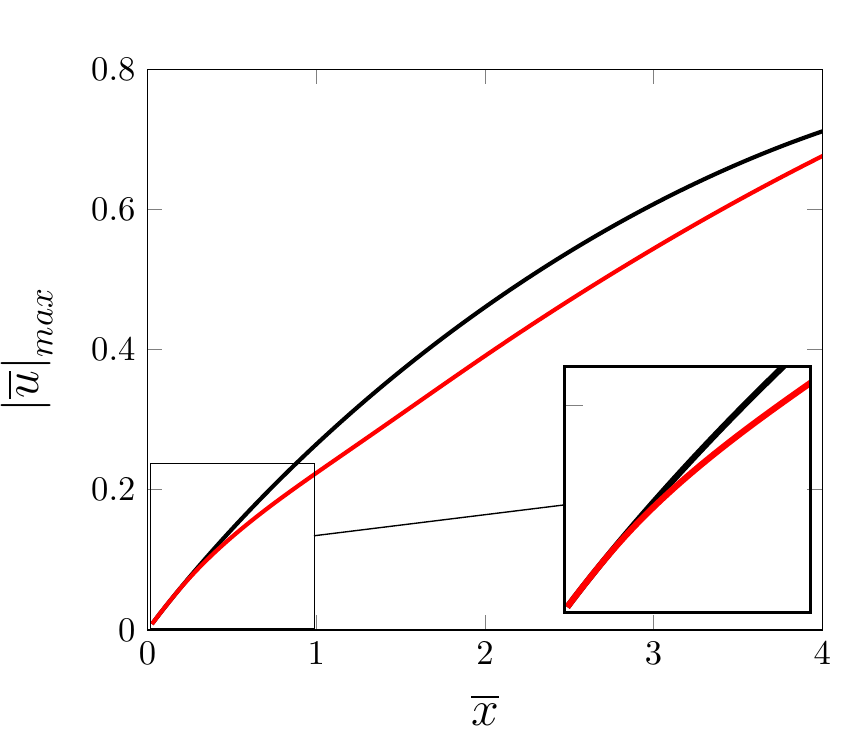}
        \caption{$\overline{x}_2=1$}
        \label{fig:effect_transition_layer_skote1}
    \end{subfigure}
    \caption{Effect of the adjustment-region length in the vicinity of the leading edge on the attenuation of the Klebanoff modes. $\overline{x}_1=0.005$ is kept constant and $\overline{x}_2$ is increased to 0.5 (figure \ref{fig:effect_transition_layer_skote05}) and 1 (figure \ref{fig:effect_transition_layer_skote1}). The solid ($\phi=0$) and porous ($\phi=0.58$) wall cases are represented by the black and red curves, respectively.}
    \label{fig:effect_transition_layer}
\end{figure}

\begin{figure}
    \centering 
    \begin{subfigure}[b]{0.49\textwidth}
        \includegraphics[width=\linewidth]{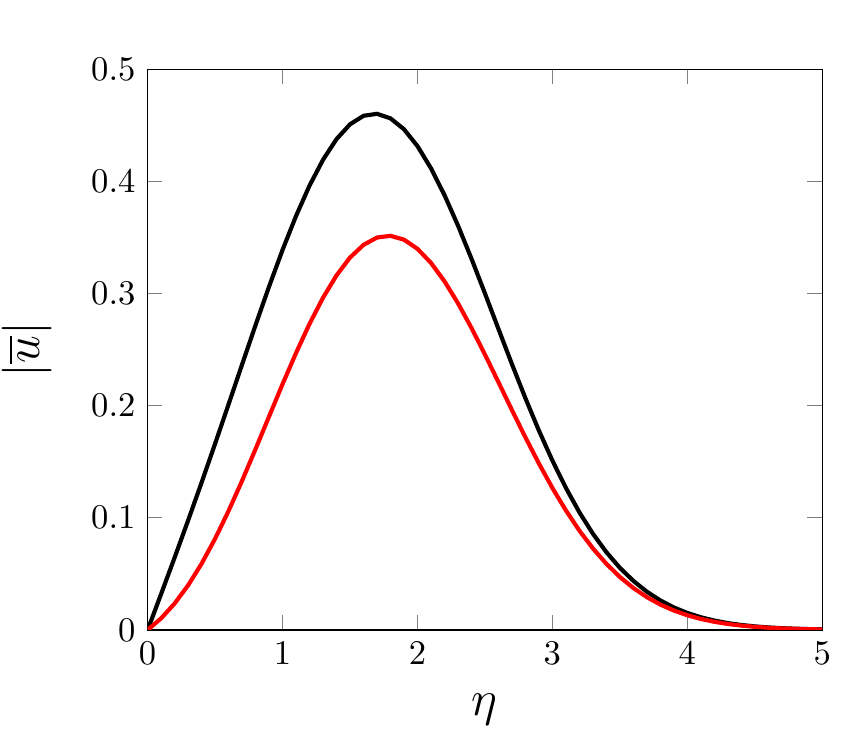}
        \caption{Streamwise velocity at $\overline{x}=2$.}\label{fig:profiles_u}
    \end{subfigure}
    \begin{subfigure}[b]{0.49\textwidth}
        \includegraphics[width=\linewidth]{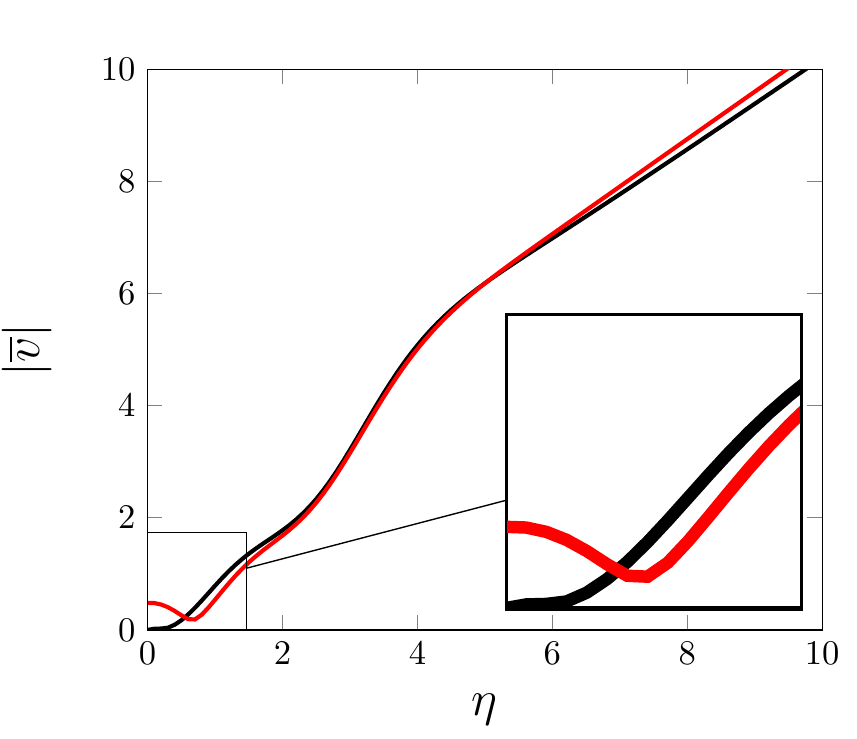}
        \caption{Wall-normal velocity at $\overline{x}=2$.}\label{fig:profiles_v}
    \end{subfigure}
    \begin{subfigure}[b]{0.49\textwidth}
        \includegraphics[width=\linewidth]{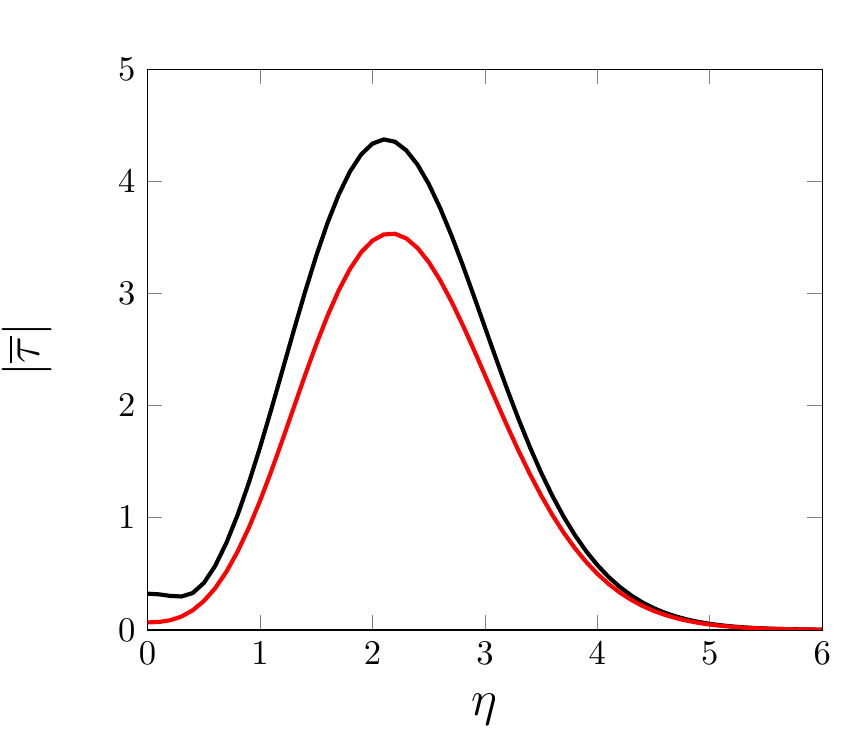}
        \caption{Temperature at $\overline{x}=2$.}\label{fig:profiles_tau}
    \end{subfigure}
    \begin{subfigure}[b]{0.49\textwidth}
        \includegraphics[width=\linewidth]{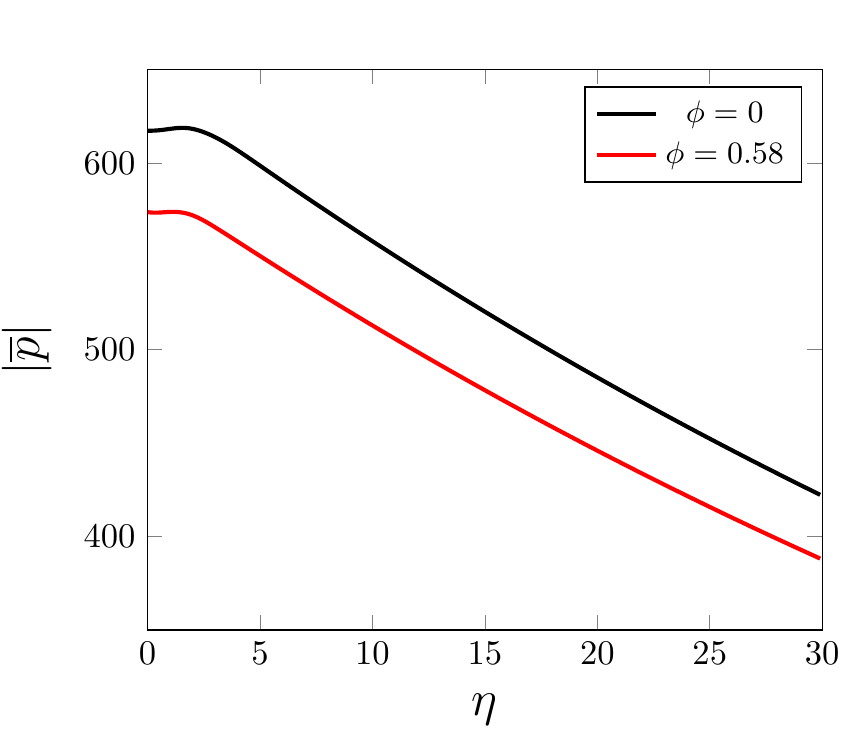}
        \caption{Pressure at $\overline{x}=2$.}\label{fig:profiles_p}
    \end{subfigure}
    \caption{Wall-normal profiles of the streamwise velocity \eqref{fig:profiles_u}, spanwise velocity \eqref{fig:profiles_v}, temperature \eqref{fig:profiles_tau}, and pressure \eqref{fig:profiles_p} disturbances at $\overline{x}=2$ for $k_z/k_x=0.3$, $\kappa_z=0.007$, and $\mathrm{R}_\lambda=418000$. The solid ($\phi=0$) and porous ($\phi=0.58$) wall cases are represented with black and red curves, respectively.} 
    \label{fig:profiles}
\end{figure}

\begin{figure}
    \begin{subfigure}[b]{0.49\textwidth}
        \includegraphics[width=\linewidth]{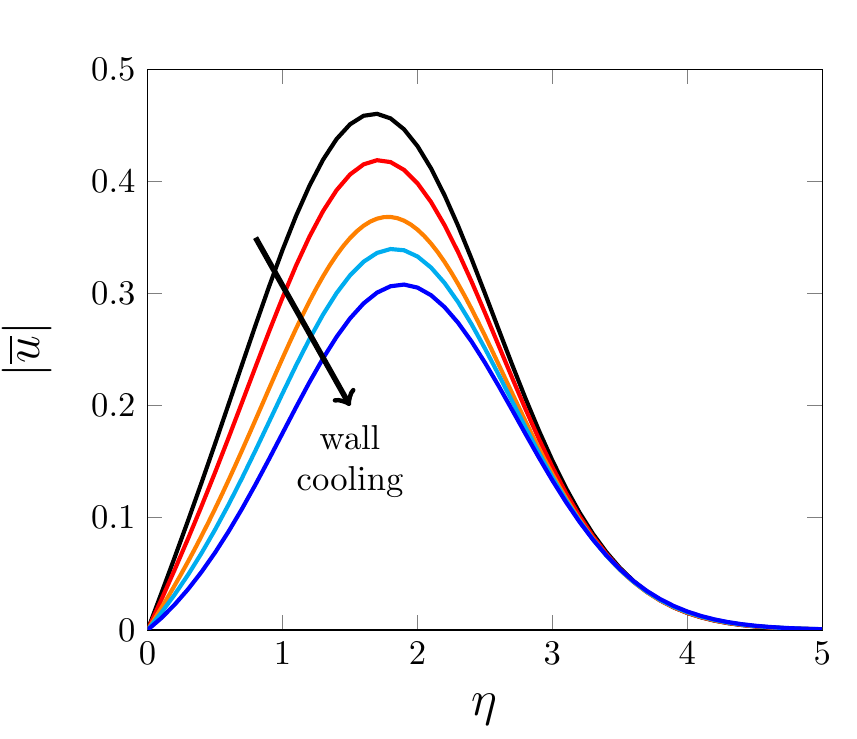}
        \caption{solid wall}\label{fig:profiles_u_wall_temperature_noporous}
    \end{subfigure}
    \begin{subfigure}[b]{0.49\textwidth}
        \includegraphics[width=\linewidth]{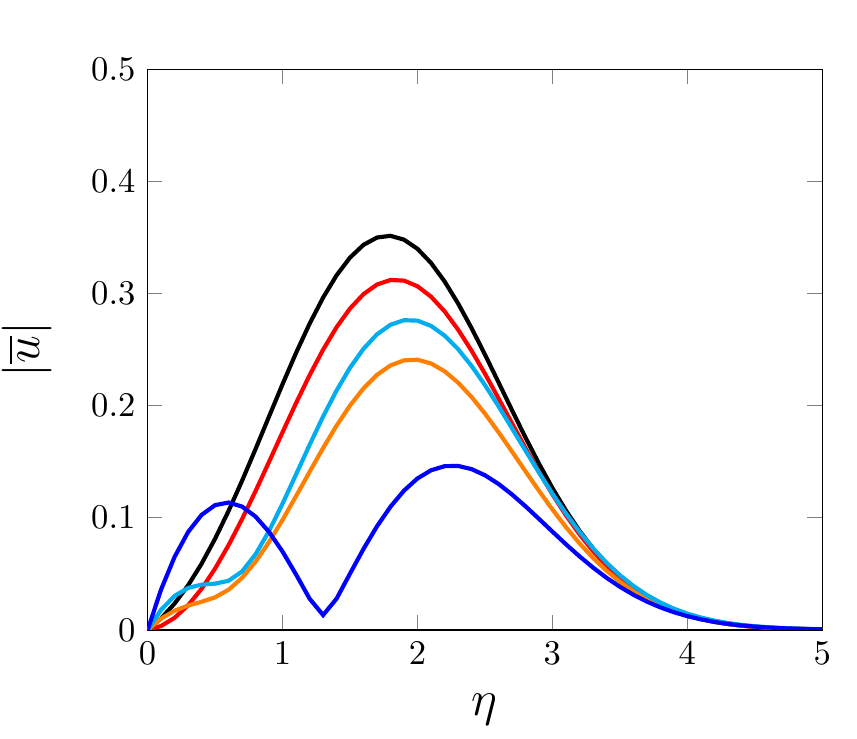}
        \caption{porous wall}\label{fig:profiles_u_wall_temperature_porous}
    \end{subfigure}
    \begin{subfigure}[b]{0.49\textwidth}
        \includegraphics[width=\linewidth]{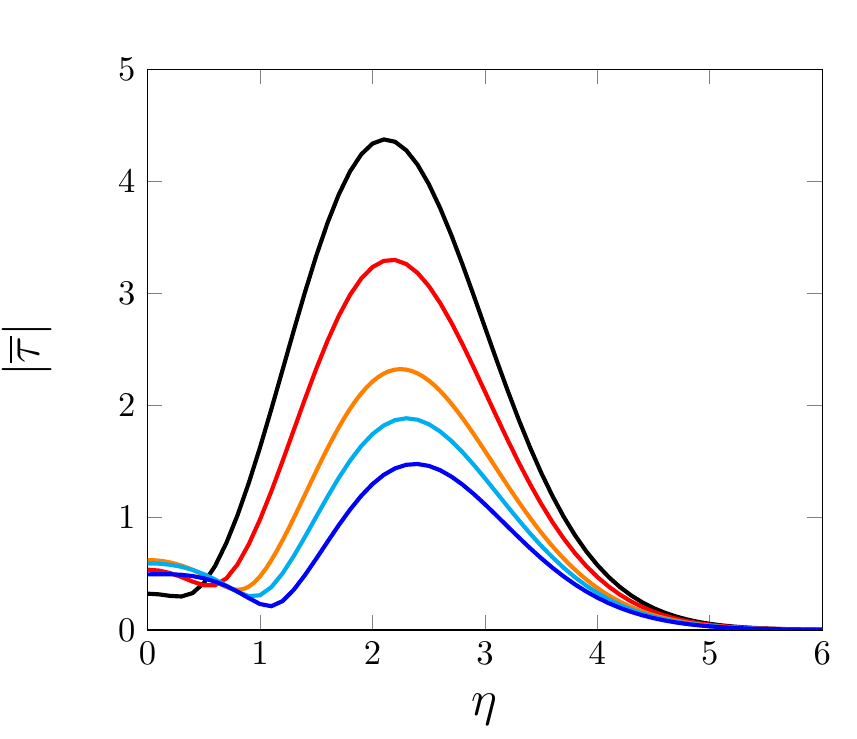}
        \caption{solid wall}\label{fig:profiles_tau_wall_temperature_noporous}
    \end{subfigure}
    \begin{subfigure}[b]{0.49\textwidth}
        \includegraphics[width=\linewidth]{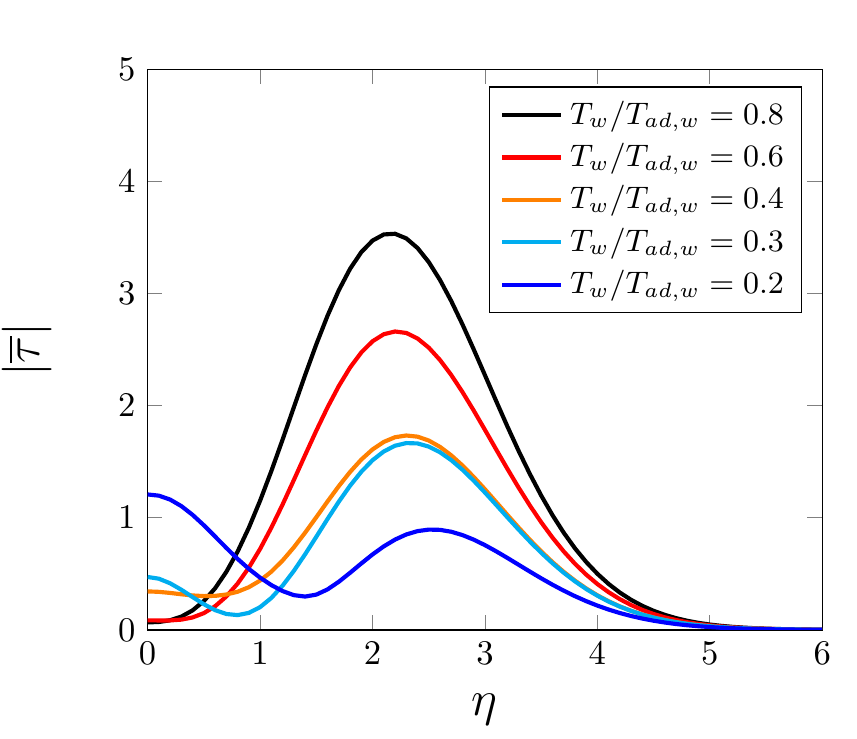}
        \caption{porous wall}
        \label{fig:profiles_tau_wall_temperature_porous}
    \end{subfigure}
    \caption{Effect of wall cooling on the streamwise velocity (top) and temperature (bottom) three-dimensional gust signatures profiles at $\overline{x}=2$ for $k_x/k_z=0.3$, $\kappa_z=0.007$, and $\mathrm{R}_\lambda=418000$. Solid and porous results are shown with solid and dashed lines, respectively.}
    \label{fig:wall_temperature}
\end{figure}

The solution of the CLUBR equations for a flat-plate boundary layer is computed for a wide range of disturbance frequencies and spanwise wavelengths. Two wall-porosity conditions are considered: a solid plate with $\phi=0$ and a porous plate with $\phi=0.58$. 

Our computations reveal that the wall porosity does not affect the growth of the Klebanoff modes for very low frequencies and very short spanwise wavelengths ($k_x\mathrm{R}_\lambda=\Or{1}$, $\kappa_z=\Or{1}$). Under these conditions, the spanwise viscous diffusion plays a significant role because $\lambda_z^\ast$ is comparable to the boundary-layer thickness $\delta^\ast$, which is typically a few millimeters \citep{Laufer_Vrebalovich_1960,Demetriades_1985,Graziosi_1999,Graziosi_Brown_2002}. The spectrum of free-stream disturbances is however wide and encompasses a wide range of spanwise wavelenghts and frequencies. We then investigate the response of the boundary layer to free-stream gusts with spanwise wavelengths that are larger than the boundary-layer thickness, i.e. $\lambda_z^\ast=0.03\,\si{\meter}$ and $\mathrm{R}_\lambda=\mathrm{R}^\ast\lambda^\ast_z=418000$. 
As $k_x\mathrm{R}_\lambda$ increases and $\kappa_z$ decreases, the effect of the porous wall becomes relevant. Its response to increasing the disturbance frequency is reported in figures \ref{fig:effect-frequency-1}, \ref{fig:effect-frequency-2}, and \ref{fig:effect-frequency-3_umax}, which show the downstream evolution of the peak of the streamwise velocity fluctuations, $\abs{\overline{u}}_{max}$, for $\kappa_z=0.008, 0.0075$ and $0.007$, respectively. Under these conditions, the scaled admittance $\overline{A}_v$ is $-8.82\cdot10^{-4}+1.434\cdot10^{-3} \ \mathrm{i}$. The growth of the Klebanoff modes is reduced by the porous wall up to about $\overline{x}=5$. The peak of the temperature fluctuations, shown in figure \ref{fig:effect-frequency-3_taumax} for $\kappa_z=0.007$, is also reduced. The attenuation becomes more significant as $\omega^\ast$ increases and $\kappa_z$ decreases, meaning that the pores absorb and dissipate the energy of the Klebanoff modes when the spanwise diffusion is small. The effectiveness of the porous layer is expected to improve at frequencies higher than those considered here. However, increasing $k_x$ beyond 0.3 might lead to a regime for which the second-order perturbation discussed in \S\ref{sec:disturbance} become important. The growth of the streamwise velocity and temperature fluctuations becomes exponential further downstream, where the receptivity of highly-oblique Tollmien-Schlichting waves sets in. This regime is studied in \S\ref{sec:TS}.

The results reported in figure \ref{fig:effect-frequency} were computed by considering the leading-edge adjustment region given by equation \eqref{skote_function} and extending between $\overline{x}_1=0.005$ and $\overline{x}_2=0.01$. The same case with $\kappa_z=0.007$ and $\mathrm{R}_\lambda=418000$ is computed for larger $\overline{x}_2$, i.e. $\overline{x}_2$=0.5 (figure \ref{fig:effect_transition_layer_skote05}) and $\overline{x}_2$=1 (figure \ref{fig:effect_transition_layer_skote1}). The growth of the Klebanoff modes is shown in figure \ref{fig:effect_transition_layer}. Extending the length of the adjustment region results in a delay of the attenuation. Albeit delayed, the damping of the Klebanoff modes is still appreciable in the region $\overline{x}\leq 4$.

More insights on the effect of wall porosity on the Klebanoff modes for $\kappa_z\ll 1$ can be inferred from the wall-normal profiles of the velocity components, the temperature and the pressure. The profiles for $\abs{\overline{u}}$, $\abs{\overline{v}}$, $\abs{\overline{\tau}}$, and $\abs{\overline{p}}$ at $\overline{x}=2$ for the case $\kappa_z=0.007$, $\mathrm{R}_\lambda=418000$ are shown in figure \ref{fig:profiles}. The streamwise velocity, the temperature, and the pressure are markedly reduced by the porous wall. The peaks of $\abs{\overline{u}}$ and $\abs{\overline{\tau}}$ are decreased and slightly shifted farther from the wall. The wall-normal gradient of $\abs{\overline{u}}$ is attenuated by the porous layer. The wall-normal velocity component $\abs{\overline{v}}$ is enhanced in the proximity of the wall (inset of figure \ref{fig:profiles_v}), but is mostly unaffected at larger wall-normal locations. The spanwise velocity component (not shown) is unchanged, which is consistent with the spanwise momentum balance being independent of $\overline{u}$, $\overline{v}$, $\overline{\tau}$ and $\overline{p}$ when $\kappa_z\ll 1$ \citepalias{Ricco_Wu_2007}. The pressure distribution retains its shape and is uniformly attenuated when the wall is porous.

The results of figure \ref{fig:effect-frequency} are computed at a fixed wall temperature ratio $T_w/T_{ad,w}=0.8$. As shown in the schematic of relation \eqref{porous_layer_effectiveness} of figure \ref{fig:porosity-regimes}, the theory indicates that a lower wall temperature increases the range of $\omega^\ast$ and $\lambda_z^\ast$ for which the pressure fluctuations are effectively transduced into wall-normal velocity fluctuations. The wall-normal profiles for the boundary-layer fluctuations at $\mathrm{M}_\infty=6$, $\kappa_z=0.007$ and $\mathrm{R}_\lambda=418000$ are computed for five different wall temperature ratios $T_w/T_{ad,w}$ and reported in figure \ref{fig:wall_temperature}. The graphs in the top row show the $\abs{\overline{u}}$-profiles over solid (figure \ref{fig:profiles_u_wall_temperature_noporous}) and porous (figure \ref{fig:profiles_u_wall_temperature_porous}) flat plates. Wall cooling uniformly reduces the amplitude of the velocity and temperature fluctuations in the solid and porous cases. 

In the solid-wall case, wall cooling causes the peak of the wall-normal profiles to shift farther from the wall, the wall-shear stress is attenuated, and the temperature fluctuations are reduced more than the velocity fluctuations, with the exception of the near-wall region where they slightly increase. The effect of wall cooling is more intense on the temperature fluctuations than on the velocity fluctuations when the wall-temperature ratio is reduced from 0.8 to 0.4. When the wall is porous, an inflection point appears close to the wall for $T_w/T_{ad,w}=0.4$, and a second shorter peak in the velocity distribution grows in the near-wall region between $T_w/T_{ad,w}=0.3$ and $T_w/T_{ad,w}=0.2$. Although the amplitude of the main velocity peak in the porous case is reduced by wall cooling, the wall-shear stress increases and the intensity of the secondary temperature peak, located at the wall, exceeds that of the main temperature peak.

\subsection{Tollmien-Schlichting waves}
\label{sec:TS}
\begin{figure}
    \centering
    \includegraphics[width=0.9\linewidth]{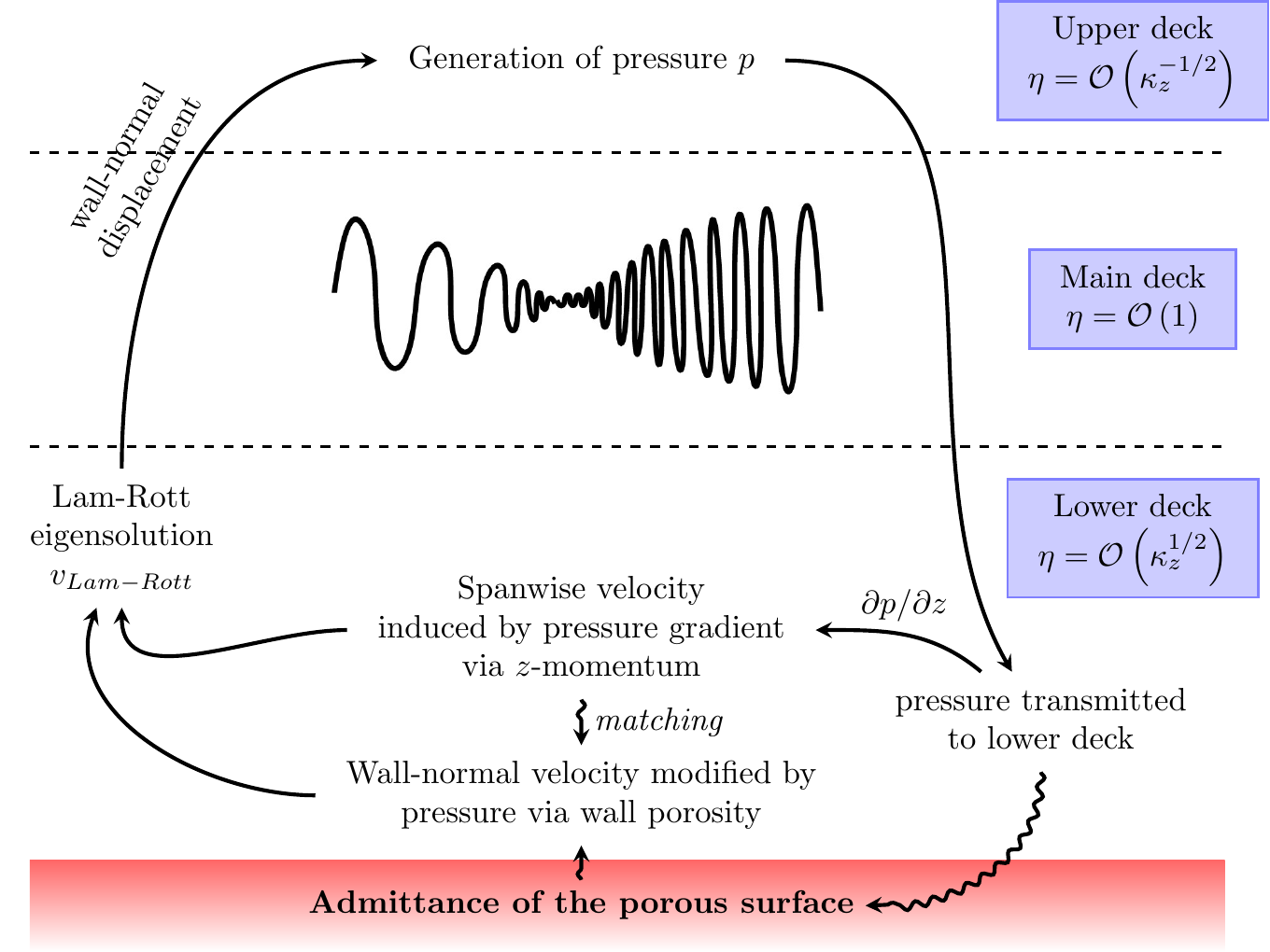}
    \caption{Schematic of triple-deck interactive regime of leading-edge receptivity mechanism in the presence of a porous surface. The thicknesses of the decks is out of scale.}
    \label{fig:triple_deck_receptivity_mechanism}
\end{figure}

For low $\kappa_z$ values, the initial algebraic growth of the compressible Klebanoff modes is followed downstream by the exponential growth of highly-oblique Tollmien-Schlichting (TS) waves, a receptivity mechanism first discovered by \citetalias{Ricco_Wu_2007}. Numerical evidence of this receptivity mechanism is shown in figure \ref{fig:effect-frequency} for $\overline{x}>6$, where the exponential growth occurs. Although the amplitude of the Klebanoff modes is attenuated, the porous wall enhances the initial amplitude of the TS waves, as also schematically illustrated in figure \ref{fig:porosity-regimes}. Our numerical result confirms the experimental findings of \cite{Shiplyuk_Burov_Maslov_Fomin_2004} and \cite{Lukashevich_Morozov_Shiplyuk_2018}. The theoretical results of \cite{Michael_Stephen_2012} also reported a larger TS-wave growth rate, although the receptivity was not included in their analysis.

The mathematical framework utilized by \citetalias{Ricco_Wu_2007} to analyze this receptivity mechanism, based on the triple-deck formalism, is extended to include wall porosity. The objectives are to verify the numerical results and to gain further insight into the modified flow instability. As the theoretical analysis is valid for $\kappa_z \ll 1$, a small value of $\kappa_z=0.0005$ is chosen for a quantitative comparison between the theoretical results and the computational data obtained by solving the CLUBR equations. 

The receptivity mechanism operates as follows. \citetalias{Ricco_Wu_2007} showed that the unsteady free-stream perturbations excite quasi three-dimensional Lam-Rott boundary-layer eigensolutions \citep{Lam_Rott_1960}, which develop downstream together with the Klebanoff modes. \cite{Goldstein_1983} first discovered that these low-amplitude decaying eigensolutions, believed until then to be innocuous for the flow instability, can turn into exponentially growing TS waves. For relatively high-frequency acoustic oscillations, \cite{Goldstein_1983} proved that the wavelength shortening of these eigensolutions indeed causes the generation of a streamwise pressure gradient that is responsible for the instability. \citetalias{Ricco_Wu_2007} instead showed that, in the low-frequency regime proper of the Klebanoff modes, a spanwise pressure gradient is induced. This pressure gradient interferes with the viscous flow by engendering a spanwise velocity component. As this component reaches the order of magnitude of the streamwise and wall-normal velocity ones, a triple-deck interacting regime sets in and a spatially-growing oblique TS wave is triggered. This receptivity mechanism is similar to the leading-edge adjustment discovered by \cite{Goldstein_1983} in that the Lam-Rott eigensolution is central for the boundary-layer dynamics. Yet, it is different because the spanwise pressure gradient is responsible for triggering the instability, while the streamwise pressure gradient is negligible. The schematic in figure \ref{fig:triple_deck_receptivity_mechanism} illustrates these physical interactions. In the case of a porous wall, the wall-normal velocity near the wall is not only altered through continuity by the spanwise velocity generated by the induced spanwise pressure gradient, but also by the wall pressure via the admittance relationship \eqref{eq:bc-vel}. The triple-deck theory has the advantage of revealing the physical mechanism responsible for engendering the first-mode growth, while this result is not achieved by performing finite-Reynolds-number stability analysis or by solving the complete Navier-Stokes equations. 

The triple-deck analysis of \citetalias{Ricco_Wu_2007} is modified to investigate how a porous surface alters the dynamics of exponentially growing unstable waves. Analogously to \citetalias{Ricco_Wu_2007}, an asymptotic eigensolution of the CLUBR equations is sought in the limits $\kappa_z \ll 1$ and $\overline{x} \gg 1$. The relevant class of eigensolutions is the one discovered by \cite{Lam_Rott_1960} (refer also to \cite{Ackerberg_Phillips_1972}). These eigensolutions are proportional to $\exp(-\hat{\psi}\overline{x}^{3/2})$, where $\hat{\psi}$ is an unknown complex eigenvalue \citep{Ackerberg_Phillips_1972,Goldstein_1983}. The eigensolutions are governed by the boundary-layer equations and the pressure disturbances need not be solved \citepalias{Leib_Wundrow_Goldstein_1999}. 
The boundary layer splits up into two decks: a main deck and a thin near-wall lower-deck. In the main deck, $\eta=\mathcal{O}(1)$ and
\begin{equation}
\label{eq:v_LR}
\left\lbrace\overline{u},\overline{v},\overline{w},\overline{\tau}\right\rbrace =
\left\lbrace\frac{F^{\prime\prime}(\eta)}{T},-\frac{3}{2}\hat{\psi}\sqrt{\overline{x}}F^\prime(\eta),0,
-\frac{T^\prime(\eta)}{T}\right\rbrace \exp(- \hat{\psi}
\overline{x}^{3/2}) + \ldots
\end{equation}
satisfy the leading-order balance in the CLUBR equations. As shown by \citetalias{Ricco_Wu_2007}, a triple-deck interactive regimes ensues because the wall-normal displacement induced downstream by the perturbation generates an unsteady pressure. This interactive regimes occurs where 
\begin{equation}
\label{eq:x_bar_TD} \overline{x} = \mathcal{O}\left(\kappa_z^{-1}\right).
\end{equation}
The decaying Lam-Rott perturbation evolves into a spatially growing, highly oblique TS wave at the locations specified by \eqref{eq:x_bar_TD} when $k_x=\order{\mathrm{R}_\lambda^{-1/5}}$ or $\kappa_z=\order{\mathrm{R}_\lambda^{-2/5}}$. As the induced pressure disturbance now plays an active role, the porosity of the wall affects the flow field. 
The streamwise coordinate 
\begin{equation}
x_1=\kappa_z \overline{x}=\mathcal{O}(1)
\label{def:x1}
\end{equation}
can be introduced because of \eqref{eq:x_bar_TD} and $\kappa_z \ll 1$. An interactive triple-deck structure emerges, consisting of a lower deck $\eta=\mathcal{O}(\kappa_z^{1/2})$, a main deck $\eta=\mathcal{O}(1)$, and an upper deck $\eta=\mathcal{O}(\kappa_z^{-1/2})$.

In the main deck, the solution expands as
\begin{equation}
\label{eq:MD_scaling}
\left\{\overline{u}, \overline{v},
\overline{w}, \overline{p},\overline{\tau}
\right\}=\left\{u_1(x_1,\eta),\kappa_z^{-1/2} v_1(x_1,\eta),
w_1(x_1,\eta), \kappa_z^{-5/2} p_1(x_1), \tau_1(x_1,\eta)\right\}E
+\ldots
\end{equation}
where

\begin{equation}
E = \exp \left( \frac{\mathrm{i}}{\kappa_z^{1/2}} \int_0^{\overline{x}} \alpha_1(x_1) \mathrm{d} \breve{x} \right).
\end{equation}
By substituting \eqref{eq:MD_scaling} into the CLUBR equations and by solving the resulting equations at leading order, one finds
\begin{equation}
\left\lbrace u_1, v_1, w_1, \tau_1\right\rbrace = \left\lbrace\mathcal{A}(x_1) F^{\prime\prime}/T, -\mathrm{i}\alpha_1
\mathcal{A}(x_1)F^\prime,p_1(x_1)T/\left(\mathrm{i}\alpha_1 F^\prime\right),-\mathcal{A}(x_1)T^\prime/T\right\rbrace,
\label{eq:MD_eq}
\end{equation}
where $\mathcal{A}(x_1)$ is an arbitrary function of $x_1$.

In the lower deck, we introduce $\underline{\eta}= \kappa_z^{-1/2} \eta = \mathcal{O}(1)$ and the leading-order solution is expressed as
\begin{equation}
\label{eq:LD_scaling}
\left\lbrace\overline{u},\overline{v},\overline{w},\overline{\tau}\right\rbrace
= \left\lbrace \overline{u}_1(x_1, \underline{\eta}), \overline{v}_1(x_1, \underline{\eta}),
\kappa_z^{-1/2} \overline{w}_1(x_1, \underline{\eta}),
\kappa_z^{1/2} \overline{\tau}_1(x_1,\underline{\eta})\right\rbrace E+\ldots. \ \
\end{equation}
Inserting \eqref{eq:LD_scaling} into the CLUBR equations yields
\begin{equation}
\left.\begin{array}{rl}
  \displaystyle \mathrm{i} \alpha_1 \overline{u}_1 +
  \frac{1}{T_w} \frac{\p \overline{v}_1}{\p
  \underline{\eta}} + \overline{w}_1 =& 0,\\
  \vspace{-.2cm}\\
  \displaystyle \mathrm{i} \left( -1 + F^{\prime\prime}(0) \alpha_1\underline{\eta}\, \right)
  \overline{u}_1 + \frac{F^{\prime\prime}(0)}{T_w} \overline{v}_1 =&
  \displaystyle  \frac{\mu_w}{2x_1T_w} \frac{\p^2
  \overline{u}_1}{\p\underline{\eta}^2},\\
  \vspace{-.2cm}\\
  \displaystyle \mathrm{i} \left( -1 + F^{\prime\prime}(0) \alpha_1\underline{\eta}\, \right)
  \overline{w}_1 = & \displaystyle T_w p_1 + \frac{\mu_w}{2x_1T_w}
  \frac{\p^2 \overline{w}_1}{\p\underline{\eta}^2}.\\
\end{array}\right\rbrace
\label{eq:LD_eq}
\end{equation}
The pressure $p_1$ in the lower deck is solely a function of $x_1$. Enforcing the no-slip condition on the streamwise and spanwise velocity components ($\overline{u}_1=0$, $\overline{w}_1=0$) in \eqref{eq:LD_eq} yields
\begin{equation}
\left.\begin{array}{rl}
  \displaystyle \left.\frac{\p \overline{v}_1}{\p \underline{\eta}}\right\vert_{\underline{\eta}=0} =& 0,\\
  \vspace{-.2cm}\\
  F^{\prime\prime}(0) \left.\overline{v}_1\right\vert_{\underline{\eta}=0} =&  \displaystyle  \frac{\mu_w}{2x_1} \left.\frac{\p^2 \overline{u}_1}{\p \underline{\eta}^2}\right\vert_{\underline{\eta}=0},\\
  \vspace{-.2cm}\\
  \displaystyle T_w p_1 + \frac{\mu_w}{2x_1T_w} \left.\frac{\p^2 \overline{w}_1}{\p \underline{\eta}^2}\right\vert_{\underline{\eta}=0} =& 0.\\
\end{array}\right\}
\label{eq:LD_eq_wall}
\end{equation}
By differentiation of the first equation in \eqref{eq:LD_eq} and by use of the second and third equations in \eqref{eq:LD_eq_wall}, one obtains
\begin{equation}
  \frac{2 \mathrm{i} x_1 \alpha_1 F^{\prime\prime}(0)}{\mu_w} \left.\overline{v}_1\right\vert_{\underline{\eta}=0} +
  \frac{1}{T_w} \left.\frac{\p^3 \overline{v}_1}{\p \underline{\eta}^3}\right\vert_{\underline{\eta}=0} -
  \frac{2 x_1 T_w^2}{\mu_w} p_1
  = 0.
\label{eq:wall}
\end{equation}
Eliminating $p_1$ from \eqref{eq:LD_eq} shows that
$\overline{v}_1$ satisfies
\begin{equation}
\left[ \frac{\p^2}{\p \underline{\eta}^2} - \frac{2\mathrm{i} x_1T_w}{\mu_w} 
\left( F^{\prime\prime}(0) \alpha_1 \underline{\eta} - 1 \right) \right] 
\frac{\p^2 \overline{v}_1}{\p \underline{\eta}^2}= 0,
\end{equation}
which has solution
\begin{equation}
\label{eq:v_sol} 
\frac{\p \overline{v}_1}{\p \underline{\eta}} = \int_{\eta_0}^{\hat
\eta} \mbox{Ai}(\breve{\eta}) \mathrm{d} \breve{\eta},
\end{equation}
where
\begin{equation}
  \widehat{\eta} = \left( 2 \mathrm{i} F^{\prime\prime}(0) \alpha_1 x_1 T_w/\mu_w
  \right)^{1/3} \underline{\eta}+\eta_0, \quad
  \eta_0 = - (\alpha_1 F^{\prime\prime}(0))^{-1}
  \left(2 \mathrm{i} F^{\prime\prime}(0) \alpha_1 x_1 T_w/\mu_w \right)^{1/3}.
\end{equation}
Differentiation of \eqref{eq:v_sol} yields 
\begin{equation}
\label{eq:v_3} 
\left.\frac{\p^3 \overline{v}_1}{\p \underline{\eta}^3}\right\vert_{\underline{\eta}=0} = 
\left( \frac{2 \mathrm{i} F^{\prime\prime}(0) \alpha_1 x_1 T_w}{\mu_w}  \right)^{2/3}
\mbox{Ai}^\prime(\eta_0).
\end{equation}
At the wall, the wall-normal velocity component and the pressure are related through \eqref{eq:bc-vel} and \eqref{eq:bc-temp}. By use of \eqref{def:x1}, \eqref{eq:MD_scaling}, and \eqref{eq:LD_scaling}, it follows that
\begin{equation}
\left.\overline{v}_1\right\vert_{\underline{\eta}=0} = 
\frac{A_v}{\kappa_z^2} \left( \frac{k_x}{ 2 x_1 R_\lambda} \right)^{1/2} \left.\overline{p}\right\vert_{\underline{\eta}=0} =
\frac{\overline{A}_v p_1}{\kappa_z^2 \left(2 x_1\right)^{1/2}}.
\end{equation}
In the case of oblique TS waves, for which $\kappa_z\ll1$ and $\overline{x}\gg1$, an admittance $A_v=\mathcal{O}\left(R_\lambda^{-1/2} k_x^{-3/2}\right)$ 
$\left(\overline{A}_v=\mathcal{O}(\kappa_z^2)\ll1 \right)$ is sufficient to alter the dynamics of the growing waves. 
A scaled admittance $\widetilde A_v=\overline{A}_v \kappa_z^{-2}=\mathcal{O}(1)$ is defined, and thus

\begin{equation}
\left.\overline{v}_1\right\vert_{\underline{\eta}=0} = \frac{\widetilde A_v p_1}{\left( 2 x_1 \right)^{1/2}}. 
\label{eq:wall_bc}
\end{equation}
The wall-normal velocity component and the pressure at the wall can now be determined. By substituting \eqref{eq:wall_bc} into \eqref{eq:wall}, it follows that
\begin{equation}
  \left( \mathrm{i} \alpha_1 F^{\prime\prime}(0) - \frac{ (2 x_1)^{1/2} T_w^2 }{\widetilde{A}_v} \right) \left.\overline{v}_1\right\vert_{\underline{\eta}=0} +
  \frac{\mu_w}{2 x_1 T_w} \left.\frac{\p^3 \overline{v}_1}{\p \underline{\eta}^3}\right\vert_{\underline{\eta}=0} 
  = 0.
\label{eq:wall_balance}
\end{equation}
By substitution of \eqref{eq:v_3} into \eqref{eq:wall_balance}, an expression for the wall-normal velocity at the wall is found
\begin{equation}
  \left.\overline{v}_1\right\vert_{\underline{\eta}=0} = \frac{\left( 2 \mathrm{i} F^{\prime\prime}(0) \alpha_1 x_1 T_w/\mu_w \right)^{2/3} \mbox{Ai}^\prime(\eta_0) \mu_w \widetilde A_v}{2 x_1 T_w \left( T_w^2 (2 x_1)^{1/2} - \mathrm{i} \widetilde A_v \alpha_1 F^{\prime\prime}(0) \right)}.
\label{eq:wall_velocity}
\end{equation}
By use of \eqref{eq:wall_bc}, the pressure in the lower deck is obtained

\begin{equation}
  \left.\overline{p}_1\right\vert_{\underline{\eta}=0} = 
  \frac{\left( 2 \mathrm{i} F^{\prime\prime}(0) \alpha_1 x_1 T_w/\mu_w \right)^{2/3} \mbox{Ai}^\prime(\eta_0) \mu_w }{(2 x_1)^{1/2} T_w \left( T_w^2 (2 x_1)^{1/2} - \mathrm{i} \widetilde{A}_v \alpha_1 F^{\prime\prime}(0) \right)}.
\label{eq:wall_pressure}
\end{equation}
Matching $\p\overline{v}_1/\p\underline{\eta}$ in \eqref{eq:v_sol} with the main-deck solution \eqref{eq:MD_eq} yields

\begin{equation}
\label{eq:disp_1} 
\int_{\eta_0}^\infty \mbox{Ai}(\breve{\eta}) \mathrm{d} \breve{\eta} = -\mathrm{i} F^{\prime\prime}(0) \alpha_1 \mathcal{A}(x_1).
\end{equation}

In the upper deck, the appropriate wall-normal variable is $\widetilde{\eta} = \kappa_z^{1/2} \eta = \mathcal{O}(1)$, and
the solution expands as

\begin{equation}
\left\lbrace\overline{u}, \overline{v},\overline{w},\overline{p},\overline{\tau}\right\rbrace
   = \left\lbrace\kappa_z^{1/2}\widetilde{u}_1(x_1,\widetilde{\eta}),
   \kappa_z^{-1/2} \widetilde{v}_1(x_1,\widetilde{\eta}),
   \widetilde{w}_1(x_1,\widetilde{\eta}), \kappa_z^{-5/2} \widetilde{p}_1(x_1,\widetilde{\eta}), 0\right\rbrace E
   +\ldots. 
   \label{eq:UD_scaling}
\end{equation}
Inserting \eqref{eq:UD_scaling} into the CLUBR equations leads to
\begin{equation} \mathrm{i} \alpha_1 \widetilde{u}_1 +  \frac{\p
\widetilde{v}_1}{\p \widetilde{\eta}} + \widetilde{w}_1 = 0,\quad
\widetilde{u}_1=0,\quad \mathrm{i} \alpha_1 \widetilde{v}_1 + \frac{1}{2 x_1}
\frac{\p \widetilde{p}_1}{\p \widetilde{\eta}} = 0,\quad \mathrm{i}
\alpha_1 \widetilde{w}_1 - \widetilde{p}_1 = 0. \label{eq:UD_eq}
\end{equation}
These equations can be reduced to a Laplace equation for $\widetilde{p}_1$,
\[
\frac{\p^2 \widetilde{p}_1}{\p \widetilde{\eta}^2} - 2 x_1 \widetilde{p}_1 = 0,
\]
whose solution is $\widetilde{p}_1 = p_1(x_1) \mbox{exp}\left(-\sqrt{2x_1}\widetilde{\eta}\right)$.
The vertical velocity behaves as $\widetilde{v}_1 \rightarrow - \mathrm{i} p_1/(\alpha_1\sqrt{2 x_1})$ for $\widetilde{\eta}
\rightarrow 0$, and matching it with the main-deck solution yields
\begin{equation}
\label{eq:disp_3} 
p_1 = \alpha_1^2 \mathcal{A}(x_1) \sqrt{2 x_1}.
\end{equation}
Eliminating $\mathcal{A}$ from \eqref{eq:disp_1} and \eqref{eq:disp_3} yields

\begin{equation}
\label{triple_deck_dispersion_relation}
\Delta(x_1,\alpha_1) \equiv \int_{\eta_0}^\infty
\mbox{Ai}(\breve{\eta}) \mathrm{d} \breve{\eta} -
\left(\frac{\mu_w}{2 \alpha_1 x_1T_w}\right)^{1/3}
\frac{\left(\mathrm{i} F^{\prime\prime}(0)\right)^{5/3} \mbox{Ai}^\prime(\eta_0)}
{ \mathrm{i} \widetilde A_v \alpha_1 F^{\prime\prime}(0) - (2 x_1)^{1/2} T_w^2 }
=0, 
\end{equation}
which is the dispersion relation that determines the complex wavenumber $\alpha_1 = \alpha_1(x_1)$. 

The admittance $\widetilde A_v$ in \eqref{triple_deck_dispersion_relation} is absent in the dispersion relation as $x_1 \rightarrow 0$, so $\mbox{Ai}^\prime(\eta_0) \rightarrow 0$ as $x_1 \rightarrow 0$. It follows from \eqref{eq:wall_velocity} that $v_1$ goes to zero at the wall as $x_1 \rightarrow 0$ and the Lam-Rott eigensolutions are therefore not influenced by the porosity at leading order.
Equation \eqref{triple_deck_dispersion_relation} reduces to the dispersion relation found by \citetalias{Ricco_Wu_2007} for a solid wall when $\widetilde A_v=0$. The Airy function and its derivative are computed by an in-house code, based on the method of \cite{Gil_Segura_Temme_2001}. The growth rate and the wavenumber are given by  $-\Im(\alpha_1)/\kappa_z^{1/2}$ and $\Re(\alpha_1)/\kappa_z^{1/2}$, respectively, and are also found numerically from the CLUBR equations as $\Re(\overline{u}_{\xbar}/\overline{u})$ and $\Im(\overline{u}_{\xbar}/\overline{u})$ (where the subscript $\overline{x}$ indicates the derivative). 

\begin{figure}
    \centering 
    \begin{subfigure}[b]{0.45\textwidth}
        \includegraphics[width=\textwidth]{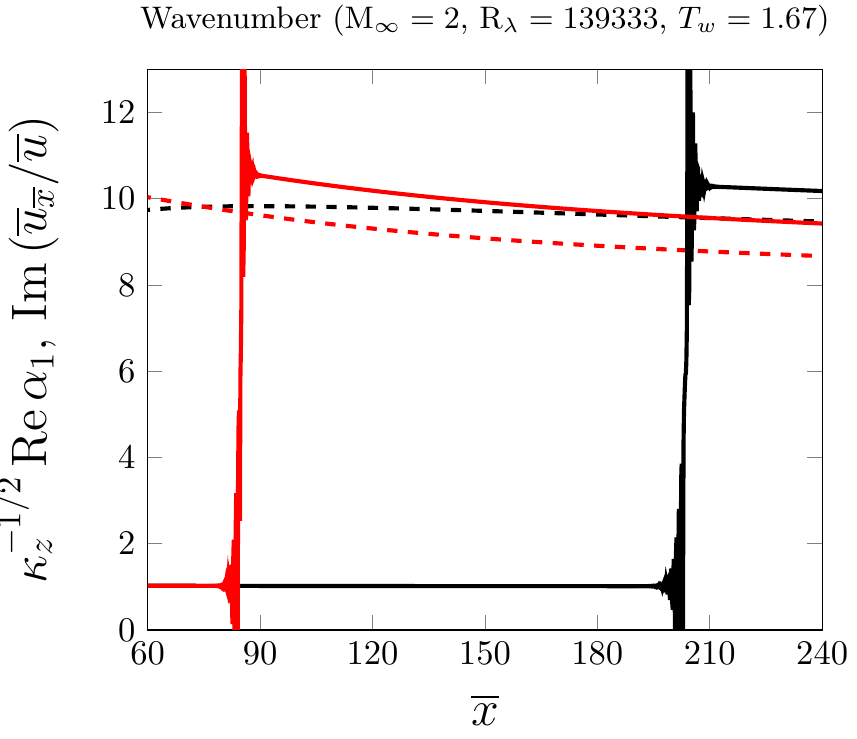}
        \label{fig:triple_deck_mach2_wavenumber}
    \end{subfigure}
    \begin{subfigure}[b]{0.45\textwidth}
        \includegraphics[width=\textwidth]{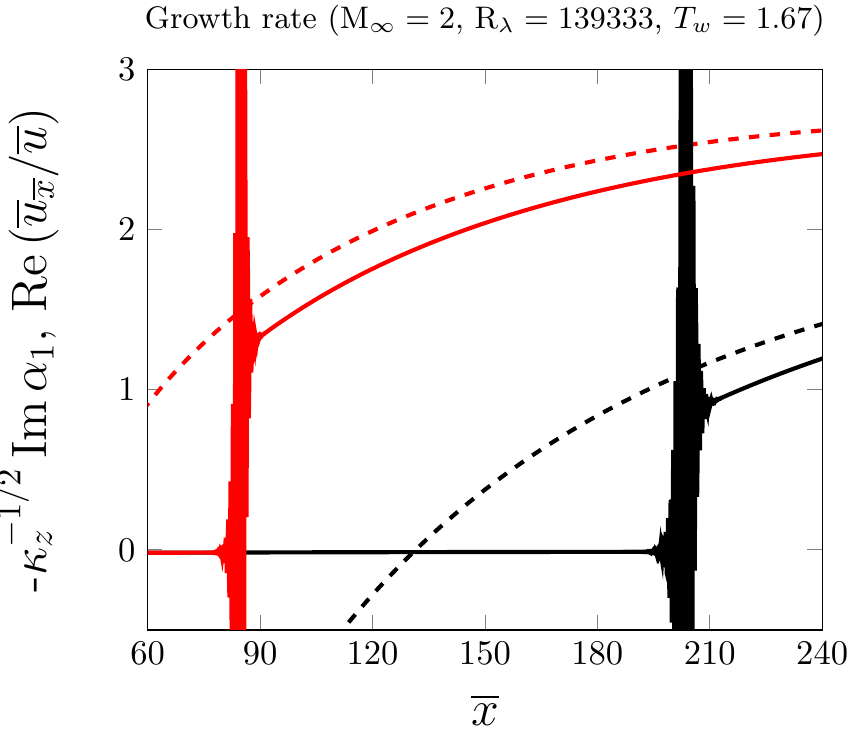}
        \label{fig:triple_deck_mach2_growth_rate}
    \end{subfigure}
    \begin{subfigure}[b]{0.45\textwidth}
        \includegraphics[width=\textwidth]{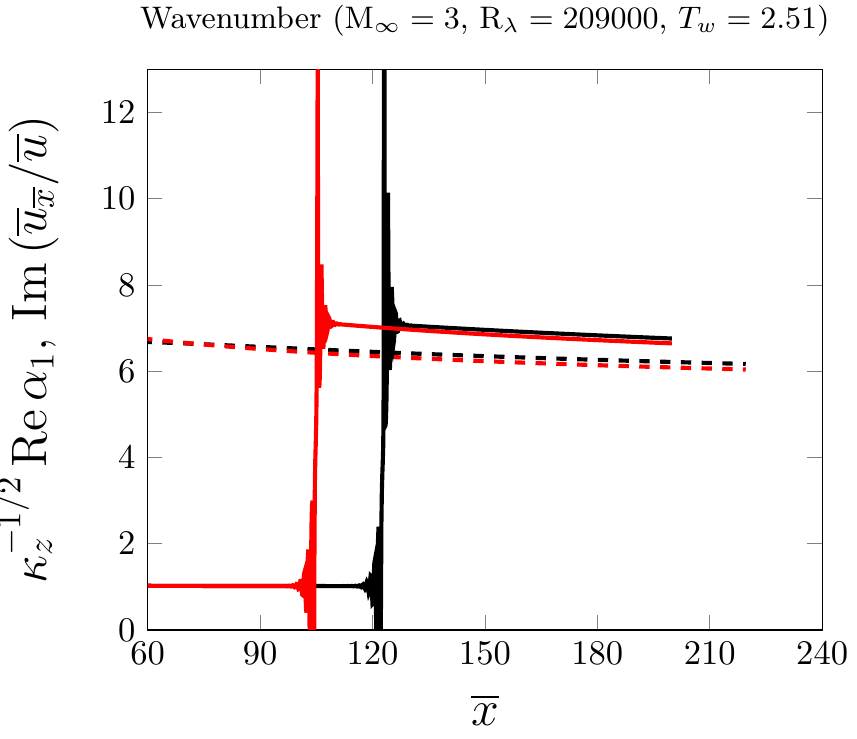}
        \label{fig:triple_deck_mach3_wavenumber}
    \end{subfigure}
    \begin{subfigure}[b]{0.45\textwidth}
        \includegraphics[width=\textwidth]{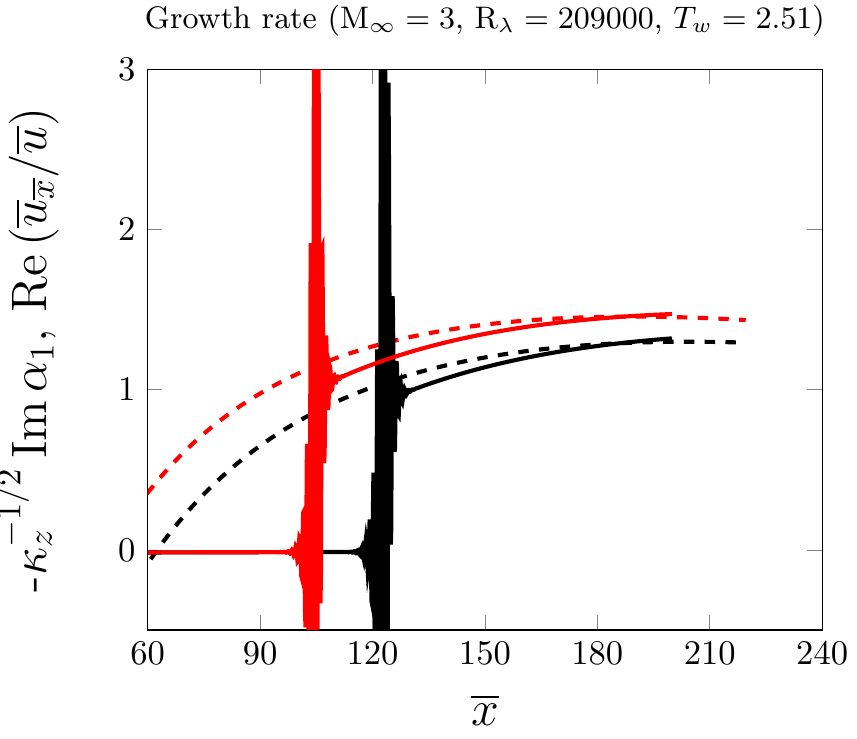}
        \label{fig:triple_deck_mach3_growth_rate}
    \end{subfigure}
    \begin{subfigure}[b]{0.45\textwidth}
        \includegraphics[width=\textwidth]{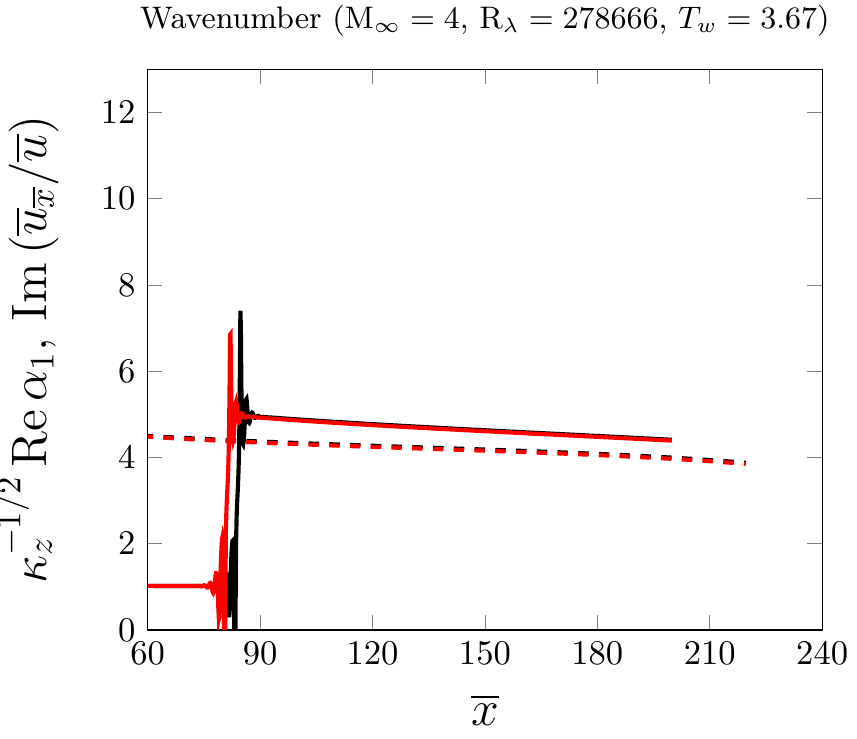}
        \label{fig:triple_deck_mach4_wavenumber}
    \end{subfigure}
    \begin{subfigure}[b]{0.45\textwidth}
        \includegraphics[width=\textwidth]{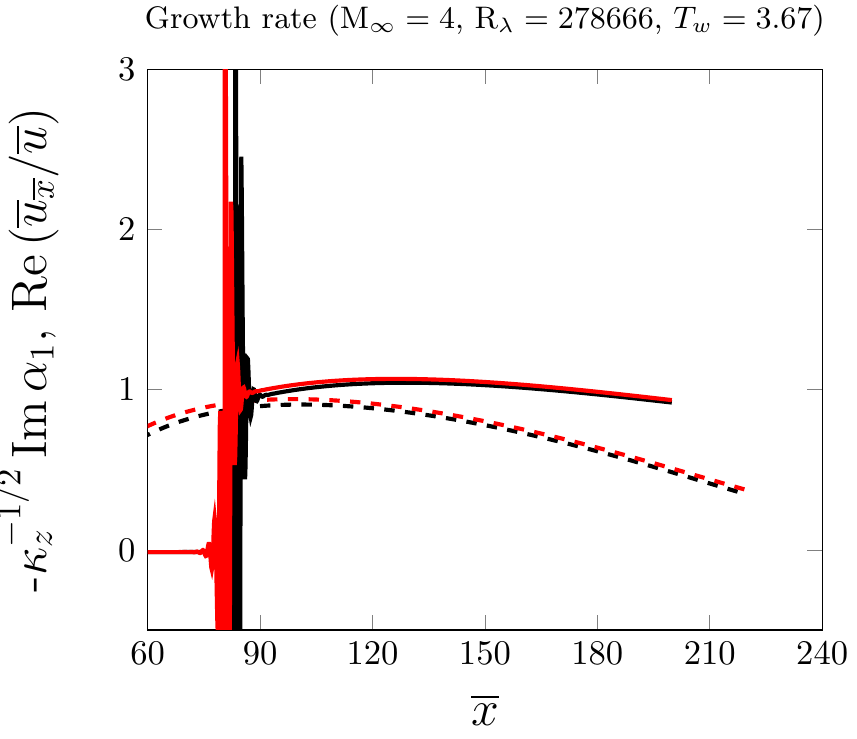}
        \label{fig:triple_deck_mach4_growth_rate}
    \end{subfigure}
    \caption{Onset of the oblique TS waves far downstream for $\kappa_z=0.0005$ on solid (black) and porous (red) walls. The solid lines indicate the boundary-region solutions and the dashed lines denote the triple-deck solutions. The three free-stream conditions at different Mach numbers are simulated by varying the free-stream velocity $U_\infty^\ast$ on adiabatic wall conditions.} 
    \label{fig:triple_deck_mach}
\end{figure}

The solutions have been first computed for $\widetilde{A}_v=\mathcal{O}(1)$ and $\mathrm{M}_\infty$=2, 3, and 4 on an adiabatic wall. The free-stream disturbances are assumed to be the same in all cases and a porous wall of fixed $R^\ast$, $\phi$ and $H^\ast$ is considered. The Mach number and Reynolds number vary together as the free-stream velocity $U_\infty^*$ increases. The wavenumber and growth rate of the CLUBR solutions and the triple-deck solutions are compared in figure \ref{fig:triple_deck_mach} for $\kappa_z=0.0005$ at $\mathrm{M}_\infty=2$, 3 and 4. The triple-deck analysis predicts the growth rate and the wavenumber of the TS instability in the solid and the porous cases, while the CLUBR solutions also give the onset of the instability. The growth rate, which is mildly negative upstream, suddenly increases as the TS waves are triggered, while the wavenumber settles to an almost constant value. The agreement between the CLUBR solutions (solid lines) and the triple-deck solutions (dashed lines) improves as the Mach number increases. The porous wall enhances the TS-wave growth rate and shifts the onset of the instability upstream, while the wavenumber is unaffected. The impact of the porous wall, however, diminishes as the Reynolds and Mach numbers increase with the free-stream velocity. For the flow conditions studied in figure \ref{fig:triple_deck_mach}, no effect of the porous wall is found at $\mathrm{M}_\infty=6$.

\begin{figure}
    {\includegraphics[width=0.49\linewidth]{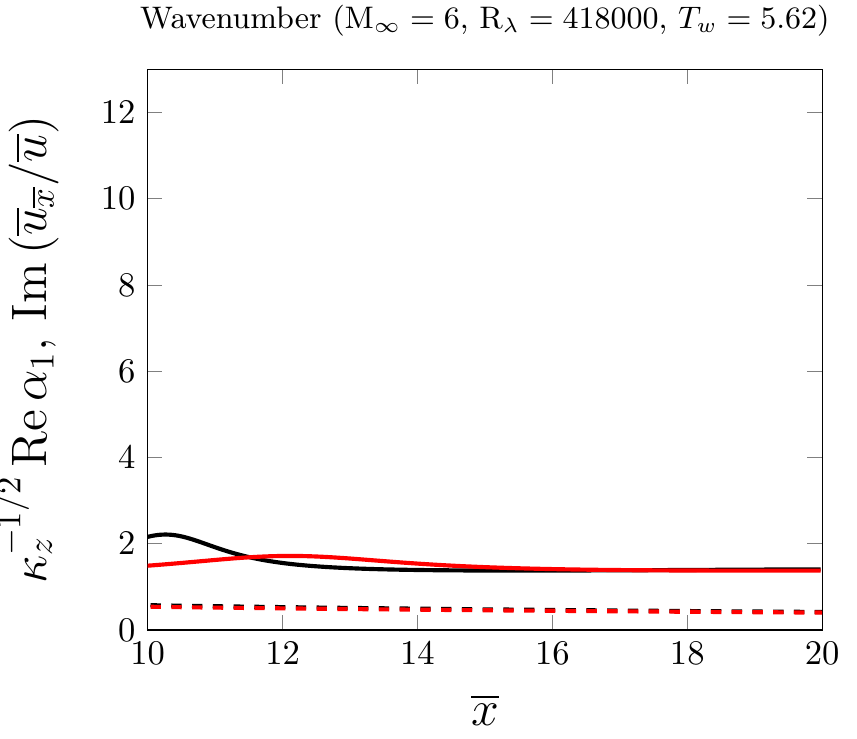}
    \includegraphics[width=0.49\linewidth]{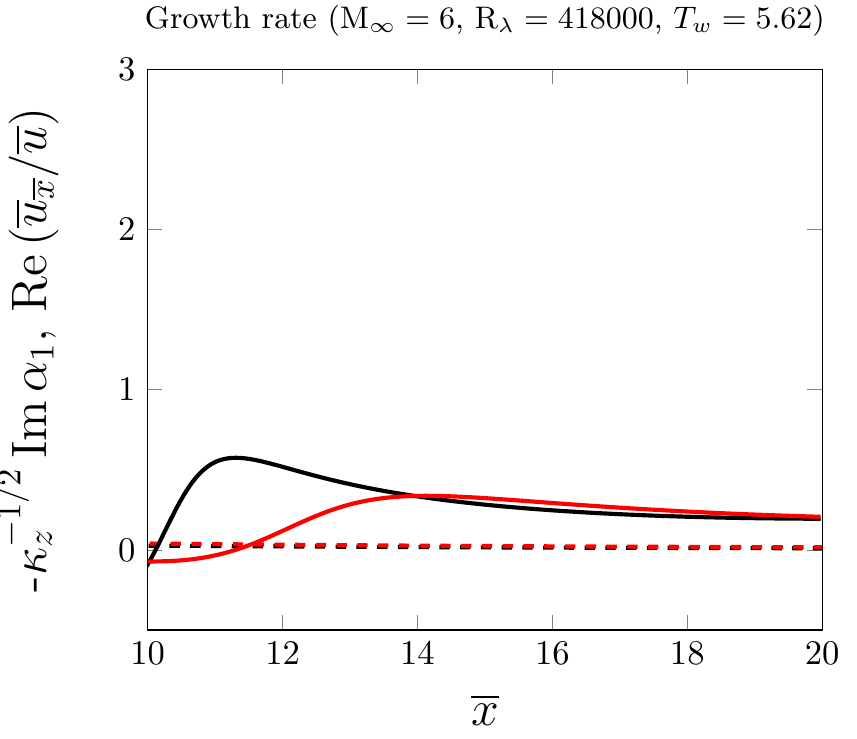}}
    \caption{Onset of the oblique TS waves for $\kappa_z=0.007$ on solid (black) and porous (red) walls. The solid lines indicate the boundary-region solutions and the dashed lines the triple-deck solutions. The results are computed for the flow conditions discussed in \S\ref{sec:k-modes}: $\mathrm{M}_\infty=6$, $T_w=5.62$ and $\kappa_z=0.007$.}
    \label{triple_deck_comparison_mach6_kappaz0007}
\end{figure}
The case investigated in \S\ref{sec:k-modes} is also studied ($\mathrm{M}_\infty=6$, $\kappa_z=0.007$, $T_w=0.8T_{ad,w}=5.62$, $\widetilde{A}_v = -18.00 +29.26 \ \mathrm{i}$). As per definition \eqref{def:x1}, the onset of the TS waves shifts upstream over both the porous and the solid surfaces when $\kappa_z$ increases slightly. The boundary-region and the triple-deck results, shown in figure \ref{triple_deck_comparison_mach6_kappaz0007}, still show a satisfactory agreement for a relatively larger $\kappa_z$ and $\overline{x}>14$. The porous wall has an intense effect on the growth rate before the exponential growth of the TS waves sets in. Once the TS-wave growth is established, the effect of porosity is mild. 

\subsection{Effect of wall curvature at moderate G\"{o}rtler number}
\label{sec:gortler}
\begin{figure}
    \centering
    \begin{subfigure}[b]{0.49\textwidth}
        \includegraphics[width=\textwidth]{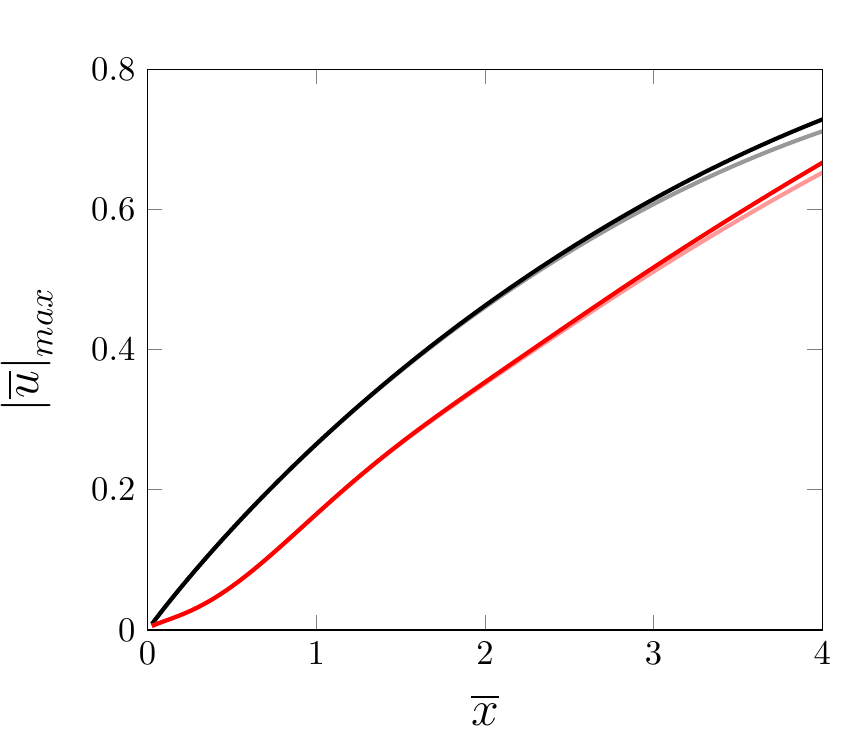}
        \caption{$\mathrm{G}=2.41$.}
        \label{fig:u_max_gortler2.41}
    \end{subfigure}
    \begin{subfigure}[b]{0.49\textwidth}
        \includegraphics[width=\textwidth]{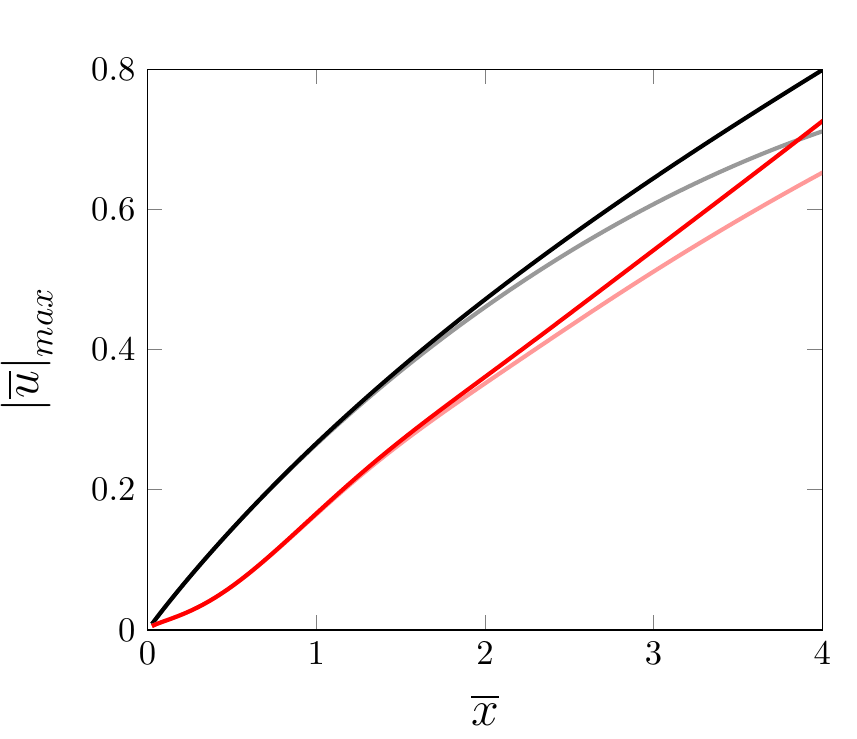}
        \caption{$\mathrm{G}=12$.}
        \label{fig:u_max_gortler12}
    \end{subfigure}
    \begin{subfigure}[b]{0.49\textwidth}
        \includegraphics[width=\textwidth]{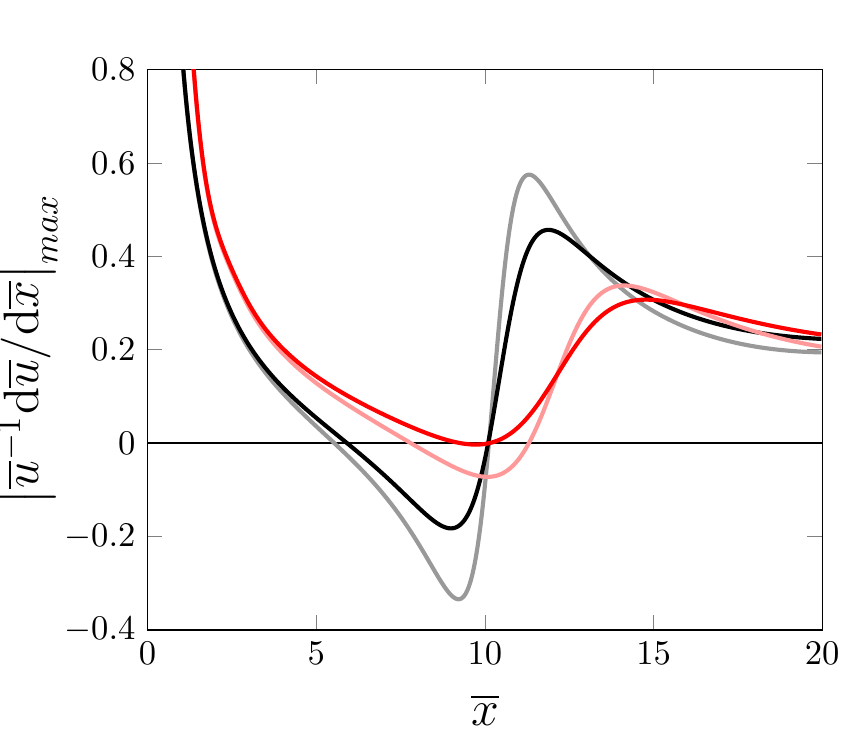}
        \caption{$\mathrm{G}=2.41$.}
        \label{fig:growth_gortler2.41}
    \end{subfigure}
    \begin{subfigure}[b]{0.49\textwidth}
        \includegraphics[width=\textwidth]{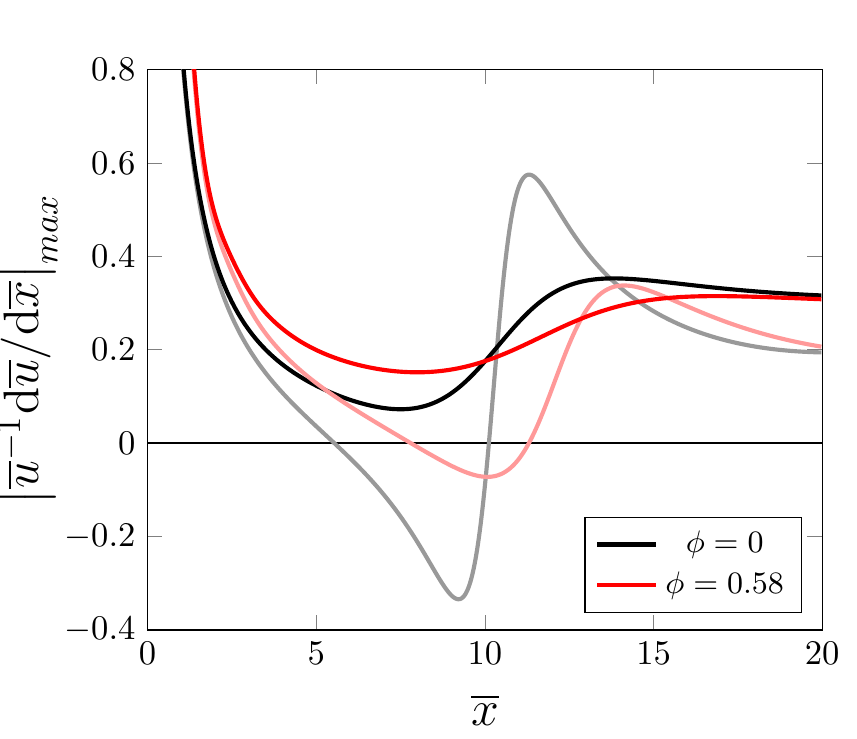}
        \caption{$\mathrm{G}=12$.}
        \label{fig:growth_gortler12}
    \end{subfigure}
    \caption{Effect of the porous wall on a boundary layer $\mathrm{G}=\order{1}$ for $\kappa_z=0.007$, $\mathrm{R}_\lambda=418000$ and $\mathrm{M}_\infty=6$. The black and red curves refer to the solid and porous cases, respectively. The curves for the flat-plate case $\mathrm{G}=0$ are drawn in light colors.}
    \label{fig:gortler}
\end{figure}

The combined effect of wall porosity and wall curvature is considered. As proved by \cite{Hall_1983}, in the limit of large Reynolds number and large curvature radius, the curvature does not affect the base flow and the centrifugal effects are distilled in two terms in the wall-normal momentum boundary-region equation \eqref{CLUBR_wall-normal_momentum} that are proportional to the G\"{o}rtler number 
\begin{equation}\label{eq:gortler_number_def}
    \mathrm{G} = \frac{1}{r_c}\left(\frac{\mathrm{R_\lambda}}{k_x^3}\right)^{1/2},
\end{equation}
where $r_c=r_c^\ast/\lambda_z^\ast$ is the scaled wall curvature radius \citep{Viaro_Ricco_2019a}. The evolution of the boundary-layer perturbations was computed for $\kappa_z=0.007$, $\mathrm{R}_\lambda=418000$, $\mathrm{M}_\infty=6$ and two different G\"{o}rtler numbers, $\mathrm{G}=2.41$ and $\mathrm{G}=12$, which correspond to $r_c=100$ and $r_c=20$, respectively. Under these conditions, the effect of curvature enhances the growth of the velocity disturbances compared to the flat-plate case. Since both $\mathrm{M}_\infty$ and $k_x\mathrm{R}_\lambda$ are relatively high, the onset of exponentially-growing G\"{o}rtler vortices was not observed \citep{Viaro_Ricco_2019b}. The downstream growth of $|\overline{u}|_{max}$ is shown in figures \ref{fig:u_max_gortler2.41} and \ref{fig:u_max_gortler12}. The flat-plate ($\mathrm{G}=0$) results are plotted in light colors for comparison. The fluctuations on the concave plate are enhanced downstream of $\overline{x}=3$ ($\mathrm{G}=2.41$) and $\overline{x}=2$ ($\mathrm{G}=12$) with respect to those on the flat plate. The porous wall reduces the amplitude of the velocity disturbances with the centrifugal effects during their initial evolution, up to about $\overline{x}=4$. Flows with higher G\"{o}rtler numbers were not investigated as values of $r_c<20$ might invalidate the hypothesis $r_c\gg\delta$.

The growth rate of $|\overline{u}|$ is shown in figures \ref{fig:growth_gortler2.41} and \ref{fig:growth_gortler12}. Although the amplitude of $\abs{\overline{u}}_{max}$ is reduced up to $\overline{x}=4$, the porous wall enhances its growth downstream of $\overline{x}=2$ up to $\overline{x}=10$ and attenuates it further downstream. Additional research is necessary to evince the effect of nonlinearity at these downstream locations because the magnitude of the fluctuations may be too large for the nonlinear interactions to be considered negligible.

\section{Conclusions}
\label{sec:conclusions}
The effect of regular-microstructure porous coatings on the receptivity of supersonic pre-transitional boundary layers to free-stream vortical disturbances has been studied. We have focused on the downstream evolution of Klebanoff modes over flat and concave surfaces and Tollmien-Schlichting waves generated by the external perturbations. We have used asymptotic and numerical methods to study these low-frequency disturbances in the limits of large Reynolds number and small amplitude. 

The downstream development of the Klebanoff modes is largely unaffected by the wall porosity when the spanwise wavelength of the oncoming perturbation is of the same order of magnitude of the boundary-layer thickness. As either the frequency or the spanwise wavelength of the disturbances increases, the wall-normal velocity and the pressure interact at the wall and the boundary-layer streamwise velocity and temperature fluctuations are attenuated. This beneficial effect is enhanced further by wall cooling.

The growth rate of Tollmien-Schlichting waves, triggered by a leading-edge adjustment mechanism, is enhanced by the wall porosity, and the location of instability moves upstream. These waves are the first modes of compressible instability, so this finding confirms previous experimental results. A triple-deck asymptotic analysis quantitatively confirms the numerical results and reveals how the physical mechanism of instability is altered in the presence of wall porosity.

The porous layer also reduces the amplitude of the streaks over concave surfaces during their initial development and the growth rate of the streamwise velocity fluctuations further downstream. 

The present work has focused on the linear response of boundary layers over porous walls to a monochromatic vortical disturbance. The full spectrum of free-stream disturbances, including acoustic and temperature fluctuations, and the nonlinear interaction of different modes should be considered in a more realistic context. Some of these aspects have been investigated by \cite{Zhang_Zaki_Sherwin_Wu_2011} and \cite{Xu_Zhang_Wu_2017,Xu_Liu_Wu_2020}. We plan to investigate the effect of wall porosity on the nonlinear compressible flows studied by \cite{Marensi_Ricco_Wu_2017} and to include higher-frequency second-mode instability in the analysis \citep{Goldstein_Ricco_2018}. Another important avenue for research is the impact of wall porosity on the secondary instability and on the final stages of transition with the objective of precisely assessing how the porous wall influences the location of transition.

Our theoretical and numerical results require further numerical and experimental validation. Direct numerical simulations and wind tunnel or in-flight measurements of the receptivity of supersonic boundary layers to free-stream vortical disturbances remain ambitious challenges. We hope our study will stimulate further research on the boundary-layer control via porous surfaces.

\section*{Acknowledgements}

The authors wish to acknowledge the support of the US Air Force through the AFOSR grant FA8655-21-1-7008 (International Program Office - Dr Douglas Smith). PR has also been supported by EPSRC (Grant No. EP/T01167X/1). The authors are also grateful to the anonymous reviewers whose comments and suggestions improved the rigour and quality of our study. PR would like to thank Professor Alexander Fedorov for the fruitful discussions and the elucidation on the acoustic field in the pores. LF would like to thank Dr Dongdong Xu and Dr Kevish Napal for all the comments, encouragement and support.

\bibliographystyle{plainnat}
\bibliography{bib}

\appendix
\section{Validation of the laminar base-flow computation}
\label{app:validation}

\begin{figure}[h]
    \centering
    {\includegraphics[width=0.49\linewidth]{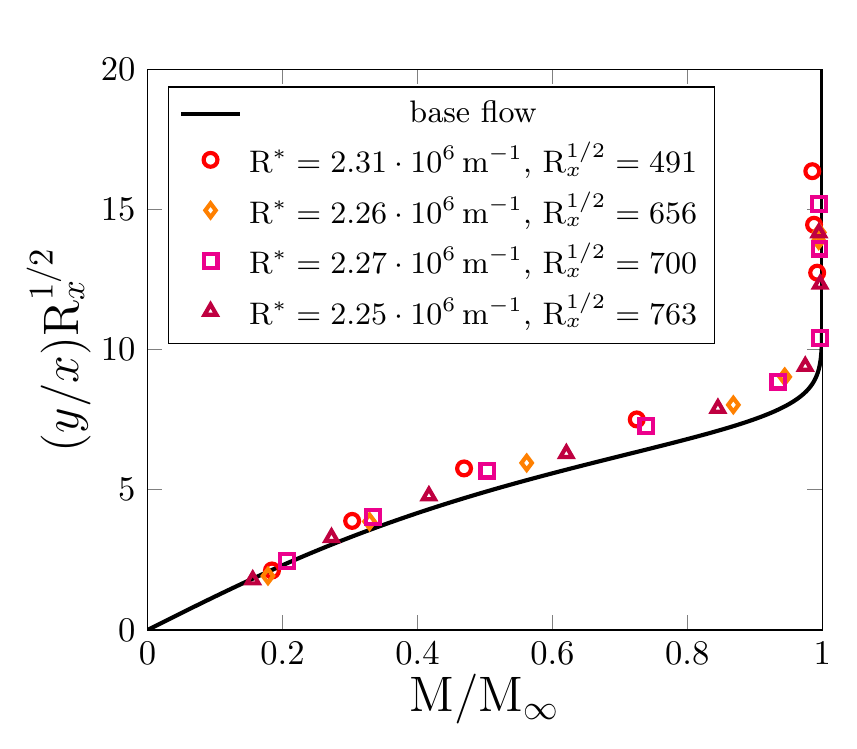}
    \includegraphics[width=0.49\linewidth]{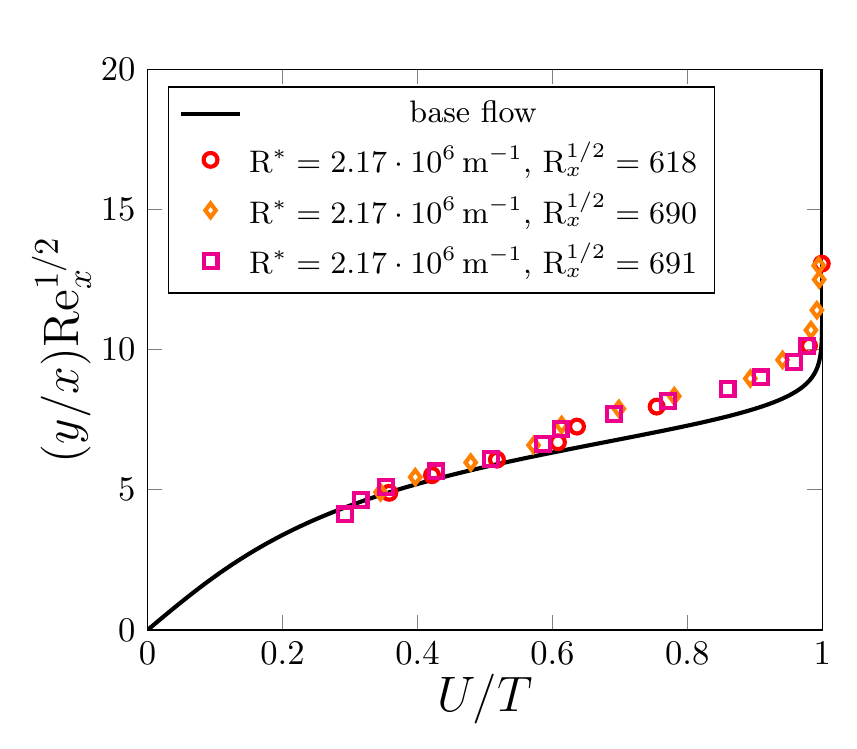}}
    \caption{Comparison of our base-flow numerical solution \eqref{base_flow_solutions} (solid lines) with the hot-wire data by \citet[pp. 85-110]{Graziosi_1999} for $\mathrm{M}_\infty=3$, $\mathrm{Pr}=0.72$, and $T_w/T_{ad,w}=1.1$.}
    \label{fig:graziosi_comparison}
\end{figure}
\begin{figure}
    \centering
    {\includegraphics[width=0.49\linewidth]{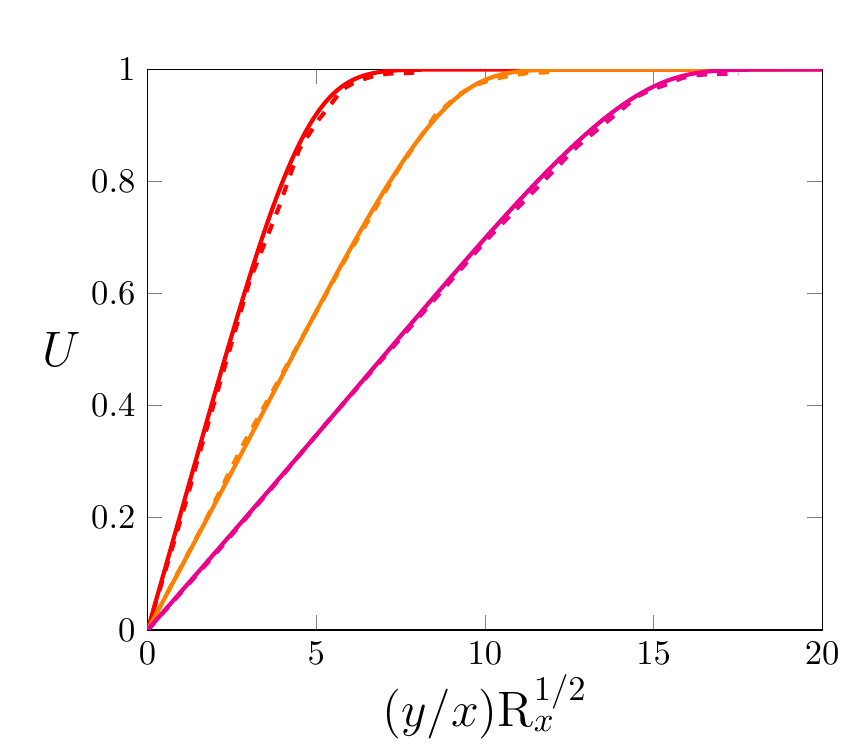}
    \includegraphics[width=0.49\linewidth]{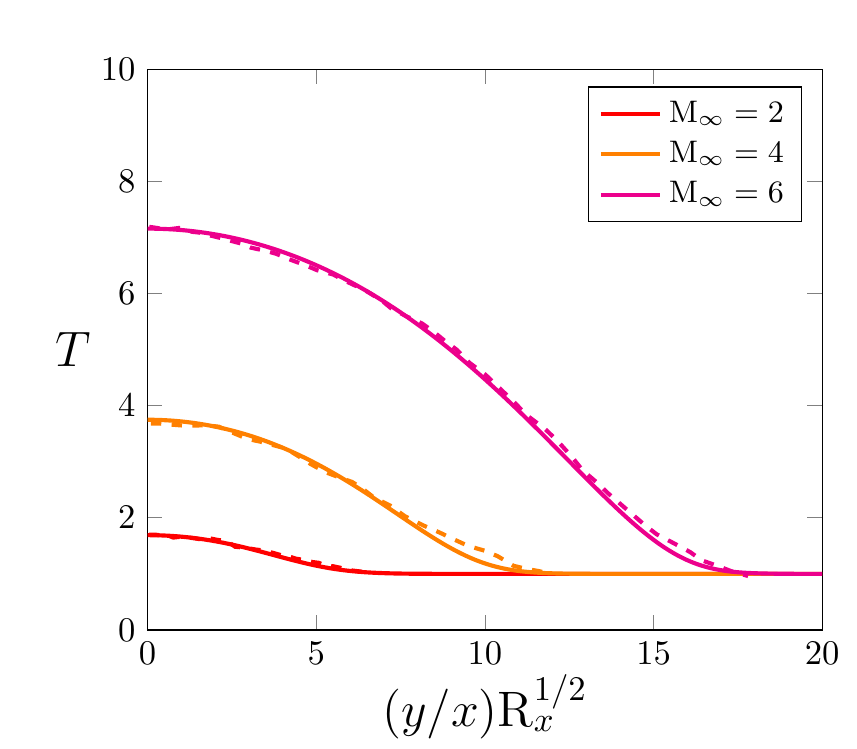}}
    \caption{Comparison of our base-flow numerical solutions \eqref{base_flow_solutions} (solid lines) with the numerical results by \citet[pp. 40-41]{vanDriest_1952} (dashed lines) for a boundary layer over an adiabatic flat plate ($\mathrm{Pr}=0.75$).}
    \label{fig:vandriest}
\end{figure}
\begin{figure}
    \centering
    \includegraphics[width=0.49\linewidth]{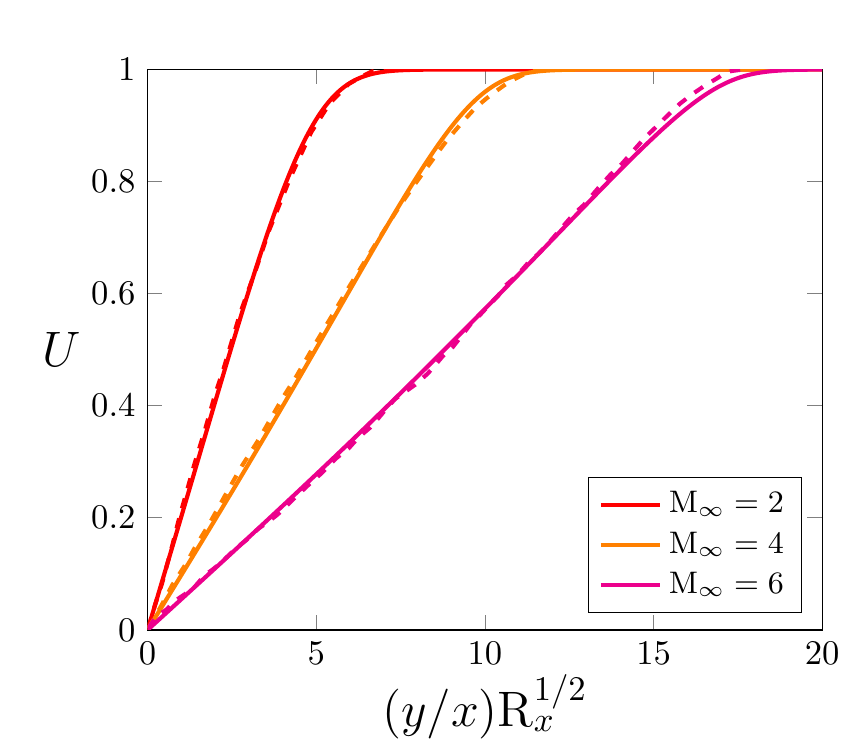}
    \caption{Comparison of our base-flow numerical solutions \eqref{base_flow_solutions} (solid lines) with the numerical results by \citet[p. 40]{Stewartson_1964} (dashed lines) for a boundary layer over an adiabatic flat plate ($\mathrm{Pr}=1.0$).}
    \label{fig:stewartson}
\end{figure}

Our numerical solutions \eqref{base_flow_solutions} of the base-flow system \eqref{compressible_Blasius} are compared to numerical and experimental data available in the literature, retrieved by the authors by using an image-digitizing software. The numerical profiles are plotted versus the similarity variable
\begin{equation}
    \widetilde{\eta} = \frac{y}{x}\mathrm{R}_x^{1/2} = 2^{1/2}\int_0^\eta T\left(\breve{\eta}\right)\mathrm{d}\breve{\eta},
\end{equation}
where $\mathrm{R}_x=U_\infty^\ast x^\ast/\nu_\infty^\ast$.

Our numerical solutions are first compared in figure \ref{fig:graziosi_comparison} with the hot-wire data by \cite[pp. 85-110]{Graziosi_1999} (refer also to \cite{Graziosi_Brown_2002}) for $\mathrm{M}_\infty=3$, $\mathrm{Pr}=0.72$, $T_w/T_{ad,w}=1.1$ and at different unit Reynolds numbers $\mathrm{R}^\ast=U_\infty^\ast/\nu_\infty^\ast$ and streamwise locations $\mathrm{R}_x^{1/2}$. Our solutions (solid lines), plotted against the local Mach number $\mathrm{M}/\mathrm{M}_\infty = U/T^{1/2}$ in figure \ref{fig:graziosi_comparison}a, show a satisfactory agreement with the experimental data.
Other experimental data at fixed $\mathrm{R}^\ast$, plotted against $U/T$ and shown in figure \ref{fig:graziosi_comparison}b, show excellent agreement for $3 \leq \widetilde{\eta} \leq 7$ and adequate agreement for $7 \leq \widetilde{\eta} \leq 10$ ($\rho T=1$ for a perfect gas has been used to convert $\rho U$ given by \cite{Graziosi_Brown_2002}). 

Our velocity and temperature profiles \eqref{base_flow_solutions} (solid lines) and those computed by \citet[pp. 40-41]{vanDriest_1952} (dashed lines) for a boundary layer over an adiabatic plate with $\mathrm{Pr}=0.75$ are shown in figure \ref{fig:vandriest}. Results were generated by modelling the dynamic viscosity with Sutherland's law $\mu = T^{3/2}(1+\chi)/(T+\chi)$, where $\chi = 0.505$, as in \citet{vanDriest_1952}. A good agreement is found for both the velocity (left) and temperature (right) profiles at all the Mach numbers. 

Figure \ref{fig:stewartson} shows that good agreement is also obtained between our solutions and the velocity profiles by \citet[p. 40]{Stewartson_1964} for a boundary layer with $\mathrm{Pr}=1$ flowing over an adiabatic plate. The dynamic viscosity was computed by using the power law $\mu=T^n$, where $n=0.76$.

\section{The compressible linearized unsteady boundary region equations}
\label{app:CLUBR}

In this appendix, the CLUBR equations, derived by \citetalias{Ricco_Wu_2007} and \cite{Viaro_Ricco_2019a}, are reported. 
\begin{itemize}
\item Continuity equation
\begin{equation}
\begin{split}
    \frac{\eta_cT^\prime}{2\overline{x}T}\overline{u} + \frac{\p\overline{u}}{\p\overline{x}} - \frac{\eta_c}{2\overline{x}}\frac{\p\overline{u}}{\p\eta} - \frac{T^\prime}{T^2}\overline{v} + \frac{1}{T}\frac{\p\overline{v}}{\p\eta} + \overline{w} &\\ 
    + \left(\frac{\mathrm{i}}{T} - \frac{FT^\prime}{2\overline{x}T^2}\right)\overline{\tau} + \frac{F^\prime}{T}
    \frac{\p\overline{\tau}}{\p\overline{x}} + \frac{F}{2\overline{x}T}\frac{\p\overline{\tau}}{\p\eta} &=0.
\end{split}
\end{equation}
\item Streamwise momentum equation
\begin{equation}
\begin{split}
    \left(-\mathrm{i} -\frac{\eta_cF^{\prime\prime}}{2\overline{x}} + \kappa_z^2\mu T\right)\overline{u} 
    + F^\prime \frac{\p\overline{u}}{\p\overline{x}} + \left(-\frac{F}{2\overline{x}} -\frac{\mu^\prime T^\prime}{2\overline{x}T} +\frac{\mu T^\prime}{2\overline{x}T^2}\right)\frac{\p\overline{u}}{\p\eta} & \\
    -\frac{\mu}{2\overline{x}T}\frac{\p^2\overline{u}}{\p\eta^2} + \frac{F^{\prime\prime}}{T}\overline{v} + 
    \left(\frac{FF^{\prime\prime}}{2\overline{x}T} - \frac{\mu^{\prime\prime}T^\prime F^{\prime\prime}}{2\overline{x}T} + \frac{\mu^\prime T^\prime F^{\prime\prime}}{2\overline{x}T^2} - \frac{\mu^\prime F^{\prime\prime\prime}}{2\overline{x}T}\right)\overline{\tau}-\frac{F^{\prime\prime}}{2\overline{x}T}\frac{\p\overline{\tau}}{\p\eta} &=0.
\end{split}
\end{equation}
\item Wall-normal momentum equation
\begin{equation}\label{CLUBR_wall-normal_momentum}
\begin{split}
    \frac{1}{\left(2\xbar\right)^2}\left(TF +\eta_c\left(FT^\prime - TF^\prime\right)-\eta_c^2 TF^{\prime\prime}\right)\utd 
    + \frac{\mu^\prime T^\prime}{3\xbar} \frac{\p\utd}{\p\xbar} &\\
    +\frac{1}{12\xbar^2}\left[\mu + \eta_cT\left(\frac{\mu}{T}\right)^\prime\right] \frac{\p\utd}{\p\eta} 
    + \frac{\eta_c\mu}{12\xbar^2} \frac{\p^2\utd}{\p\eta^2} & \\
    -\frac{\mu}{6\xbar} \frac{\p^2 \utd}{\p\xbar \p\eta} + 
    \left(-\mathrm{i} + \frac{F^\prime}{2\xbar} -\frac{T^\prime F}{2\xbar T} +\frac{\eta_c F^{\prime\prime}}{2\xbar} + \kappa_z^2 \mu T\right)\vtd 
    + F^\prime \frac{\p\vtd}{\p\xbar} & \\
    + \left(-\frac{F}{2\xbar} - \frac{2\mu^\prime T^\prime}{3\xbar T} + \frac{2\mu T^\prime}{3\xbar T^2} \right)\frac{\p\vtd}{\p\eta} -\frac{2\mu }{3\xbar T}\frac{\p^2\vtd}{\p\eta^2} + 
    \frac{\mu^\prime T^\prime}{3\xbar} \wtd -\frac{\mu}{6\xbar} \frac{\p\wtd}{\p\eta} & \\
    +\frac{1}{\left(2\xbar \right)^2}\left[\eta_c\left(FF^\prime\right)^\prime-FF^\prime -\frac{T^\prime F^2}{T} -\mu^\prime F^{\prime\prime} -\eta_c T\left(\frac{\mu^\prime F^{\prime\prime}}{T}\right)^\prime 
    +\frac{4}{3}\left(\frac{\mu^\prime T^\prime F}{T}\right)^\prime \right] \tautd & \\
    -\frac{\mu^\prime F^{\prime\prime}}{2\xbar} \frac{\p\tautd}{\p\xbar} + 
    \left[-\frac{\mu^\prime\eta_c F^{\prime\prime}}{\left(2\xbar\right)^2} +\frac{4}{3}\frac{\mu^\prime T^\prime F}{\left(2\xbar\right)^2T} \right] \frac{\p\tautd}{\p\eta} + \frac{1}{2\xbar} \frac{\p\ptd}{\p\eta} & \\
    +\boxed{\frac{\mathrm{G}}{\sqrt{2\xbar}}\left(2F^\prime\utd - \frac{\left(F^\prime\right)^2}{T}\tautd\right)}&=0,
\end{split}
\end{equation}
where the terms that distill the effect of the curvature are enclosed in the box \citep{Viaro_Ricco_2019a} and the G\"{o}rtler number $\mathrm{G}$ is defined in equation \eqref{eq:gortler_number_def}. 
\item Spanwise momentum equation
\begin{equation}
\begin{split}
    -\kappa_z^2\frac{\eta_c\mu^\prime T T^\prime}{2\xbar}\utd + 
    \kappa_z^2\frac{\mu T}{3}\frac{\p\utd}{\p\xbar} - \kappa_z^2\frac{\eta_c\mu T}{6\xbar} \frac{\p\utd}{\p\eta} + \kappa_z^2\mu^\prime T^\prime\vtd + \kappa_z^2\frac{\mu}{3} \frac{\p\vtd}{\p\eta}  & \\
    + \left(-\mathrm{i} + \frac{4}{3}\kappa_z^2\mu T \right)\wtd+ F^\prime \frac{\p\wtd}{\p\xbar} - \left(\frac{F}{2\xbar} + \frac{\mu^\prime T^\prime}{2\xbar T} - \frac{\mu T^\prime}{2\xbar T^2}\right)\frac{\p\wtd}{\p\eta} - \frac{\mu}{2\xbar T}\frac{\p^2\wtd}{\p\eta^2}  & \\
    - \kappa_z^2T\ptd +\kappa_z^2\frac{\mu^\prime T^\prime F}{3\xbar}\tautd &=0.
\end{split}
\end{equation}
\item Energy equation
\begin{equation}
\begin{split}
    -\frac{\eta_cT^\prime}{2\overline{x}}\overline{u} + \left(\gamma-1\right)\mathrm{M}_\infty^2\frac{\mu F^{\prime\prime}}{\overline{x}T}\frac{\p\overline{u}}{\p\eta} + \frac{T^\prime}{T}\overline{v} &\\
    +\left[-\mathrm{i} + \frac{FT^\prime}{2\overline{x}T} - \left(\gamma-1\right)\mathrm{M}_\infty^2\frac{\mu^\prime\left(F^{\prime\prime}\right)^2}{2\overline{x}T} - \frac{1}{2\overline{x}\mathrm{Pr}}\left(\frac{\mu^\prime T^\prime}{T}\right)^\prime + \frac{\mu\kappa_z^2T}{\mathrm{Pr}}\right]\overline{\tau} & \\
    +F^\prime \frac{\p\overline{\tau}}{\p\overline{x}} + 
    \left(- \frac{F}{2\overline{x}} - \frac{1}{\mathrm{Pr}}\frac{\mu^\prime T^\prime}{2\overline{x}T} + \frac{1}{\mathrm{Pr}}\frac{\mu T^\prime}{2\overline{x} T^2}\right)\frac{\p\overline{\tau}}{\p\eta} 
    + \frac{1}{\mathrm{Pr}}\frac{\mu}{2\overline{x}T}\frac{\p^2\overline{\tau}}{\p\eta^2} &=0.
\end{split}
\end{equation}
\end{itemize}

\section{Admittances of the porous wall}
\label{app:appendix_admittance}

We consider a single pore oriented along the wall-normal direction $y$ and located underneath the wall \citep{Zwikker_Kosten_1949,Biot_1956-prop-theory-part2,Stinson_1991,Fedorov_Malmuth_Rasheed_Hornung_2001}. The depth of the pore $H^\ast$ is much larger than its radius $R^\ast$, and the propagation of the disturbances is described in a cylindrical coordinate system. Since the pore is long and thin, and the average velocity is zero therein, one can assume the radial and azimuthal components of the velocity disturbance to be zero. The dynamic viscosity and the thermal conductivity are assumed constant, as the perturbations of the temperature field are small in amplitude. The axial coordinate is scaled with $H^\ast$ and the radial coordinate is scaled with $R^\ast$. The time is scaled by the angular frequency $\omega^\ast$ and the pressure is scaled by $\rho_w^\ast\left(H^\ast\omega^\ast\right)^2$, where the density $\rho^\ast_w$ is the density at the boundary-layer interface. The scaled quantities are denoted by the superscript $\bullet$. The pore has an open end at $y^\bullet=0$ and is closed at $y^\bullet=-1$. 

Since $r^\ast/H^\ast\gg 1$ one can assume the pressure disturbance to propagate as a planar wave along the pore \citep{Kinsler_Frey_Coppens_Sanders_2000,Stinson_1991}. Harmonic disturbances of the type
\begin{subequations}
\begin{align}
    p^\bullet\left(y^\bullet;t^\bullet\right) &= \widetilde{p}^\bullet\left(y^\bullet\right)e^{-\mathrm{i} t^\bullet} \\
    v^\bullet\left(r^\bullet;y^\bullet;t^\bullet\right) &= \widetilde{v}^\bullet\left(r^\bullet;y^\bullet\right)e^{-\mathrm{i} t^\bullet} \\
    \tau^\bullet\left(r^\bullet;y^\bullet;t^\bullet\right) &= \widetilde{\tau}^\bullet\left(r^\bullet;y^\bullet\right)e^{-\mathrm{i} t^\bullet}
\end{align}
\end{subequations}
are introduced in the continuity equation, the axial momentum and energy equations, and the perfect gas equation, which take the linearized form 
\begin{subequations}
\label{REG_POROUS_cylindrical}
\begin{align}
    -\mathrm{i}\widetilde{\rho}^\bullet + \frac{\p\widetilde{v}^\bullet}{\p y^\bullet} &=0, \\
    -\mathrm{i}\widetilde{v}^\bullet + \frac{\mathrm{d}\widetilde{p}^\bullet}{\mathrm{d} y^\bullet} &= \frac{1}{K_v^2}\left(\frac{\p^2\widetilde{v}^\bullet}{\p r^{\bullet2}} + \frac{1}{r^\bullet}\frac{\p\widetilde{v}^\bullet}{\p r^\bullet}\right), \label{REG_POROUS_cylindrical_momentum_balance}\\
    -\widetilde{\tau}^\bullet &= -\left(\gamma-1\right)\mathrm{He}^2\widetilde{p}^\bullet + \frac{1}{\mathrm{Pr}}\frac{1}{\mathrm{i} k_v^2}\left(\frac{\p^2\widetilde{\tau}^\bullet}{\p r^{\bullet2}} + \frac{1}{r^\bullet}\frac{\p\widetilde{\tau}^\bullet}{\p r^\bullet}\right), \label{REG_POROUS_cylindrical_energy_balance}\\
    \gamma\mathrm{He}^2\widetilde{p}^\bullet &= \widetilde{\rho}^\bullet + \widetilde{\tau}^\bullet, \label{REG_POROUS_perfect_gas}
\end{align}
\end{subequations}
where $\mathrm{He}=\omega^\ast H^\ast/c_w^\ast = \mathcal{O}(1)$ is the Helmholtz number of the pore and $K_v$ is defined in \eqref{Kv_parameter}.

The solutions that satisfy no-slip and isothermal boundary conditions at the wall are
\begin{subequations}\label{REG_POROUS_cylindrical_solutions}
\begin{align}
    \widetilde{v}^\bullet\left(r^\bullet,y^\bullet\right) &= 
    -\mathrm{i}\frac{\mathrm{d}\widetilde{p}^\bullet}{\mathrm{d} y^\bullet}\left[1-\frac{J_0\left(\mathrm{i}^{1/2}K_vr^\bullet\right)}{J_0\left(\mathrm{i}^{1/2}K_v\right)}\right], \label{REG_POROUS_cylindrical_solutions_velocity} \\
    \widetilde{\tau}^\bullet\left(r^\bullet,y^\bullet\right) &= \left(\gamma-1\right)\mathrm{He}^2\widetilde{p}^\bullet\left[1-\frac{J_0\left(\left(\mathrm{i}\mathrm{Pr}\right)^{1/2}K_v r^\bullet \right)}{J_0\left(\left(\mathrm{i}\mathrm{Pr}\right)^{1/2}\right)}\right],
\end{align}
\end{subequations}
where $J_0$ and $J_1$ are the Bessel functions of the first kind of order 0 and 1, respectively.

The cross-sectional averages of the velocity and temperature solutions are 
\begin{subequations}
\label{REG_POROUS_cylindrical_averages}
\begin{align}
    \average{\widetilde{v}^\bullet}\left(y^\bullet\right) &= -\mathrm{i}\frac{\mathrm{d}\widetilde{p}^\bullet}{\mathrm{d} y^\bullet}\left[1-\mathcal{F}\left(\mathrm{i}^{1/2}K_v\right)\right], \label{REG_POROUS_cylindrical_averages_velocity} \\
    \average{\widetilde{\tau}^\bullet}\left(y^\bullet\right) &= \left(\gamma-1\right)\mathrm{He}^2\widetilde{p}^\bullet\left[1-\mathcal{F}\left(\left(\mathrm{i}\mathrm{Pr}\right)^{1/2}K_v\right)\right], \label{REG_POROUS_cylindrical_averages_temperature}
\end{align}
\end{subequations}
and $\mathcal{F}$ is a complex function, 
\begin{equation}
\label{eq:F-hat}
    \mathcal{F}\left(\xi\right) = \frac{2J_1\left(\xi\right)}{\xi J_0\left(\xi\right)} = 1+ \frac{J_2\left(\xi\right)}{J_0\left(\xi\right)}.
\end{equation}
The pressure disturbance satisfies the equation \citep{Stinson_1991}
\begin{equation}
\label{REG_POROUS_helmholtz}
    \frac{\mathrm{d}^2\widetilde{p}^\bullet}{\mathrm{d} y^{\bullet2}} - \Lambda^2\widetilde{p}^\bullet = 0.
\end{equation}
The solution to \eqref{REG_POROUS_helmholtz} is
\begin{equation}
    \widetilde{p}^\bullet\left(y^\bullet\right) = a\left[e^{-\Lambda^\bullet\left(y^\bullet+1\right)}+ e^{\Lambda^\bullet\left(y^\bullet+1\right)}\right],
\end{equation}
where $a$ is a real constant and $\Lambda$ is the propagation constant defined as
\begin{equation}
\label{REG_POROUS_prop_constant}
    \Lambda^\bullet=\mathrm{i}\left(Z_1^\bullet Y_1^\bullet\right)^{1/2},
\end{equation}
$Z_1^\bullet$ and $Y_1^\bullet$ are the non-dimensional series impedance (dynamic density) and shunt admittance (dynamic compressibility)

\begin{subequations}
\begin{align}
    Z_1^\bullet &= \left[1-\mathcal{F}\left(\mathrm{i}^{1/2}K_v\right)\right]^{-1}, \\
    Y_1^\bullet &= \mathrm{He}^2 \left[1 +\left(\gamma-1\right)\mathcal{F}\left(\left(\mathrm{i}\mathrm{Pr}\right)^{1/2}K_v\right)\right].
\end{align}
\end{subequations}

The velocity and temperature admittances at the pore inlet ($y^\ast=0$) is given by the ratios of the velocity and temperature to pressure

\begin{subequations}
\label{REG_POROUS_admittance}
\begin{align}
    A_v^\bullet\left(0\right) =\frac{\average{\widetilde{v}^\bullet}\left(0\right)}{\widetilde{p}^\bullet\left(0\right)} &= -\frac{\Lambda^\bullet}{\mathrm{i} Z_1^\bullet} \tanh{\left(\Lambda^\bullet\right)}, \label{REG_POROUS_admittance_fedorov_form} \\
    A_\tau^\bullet\left(0\right) = \frac{\average{\widetilde{\tau}^\bullet}\left(0\right)}{\widetilde{p}^\bullet\left(0\right)} &= \left(\gamma-1\right)\mathrm{He}^2\left[1-\mathcal{F}\left(\left(\mathrm{i}\mathrm{Pr}\right)^{1/2}K_v\right)\right]. \label{REG_POROUS_admittance_temperature}
\end{align}
\end{subequations}
The velocity admittance \eqref{REG_POROUS_admittance_fedorov_form} is expressed by means of either the propagation constant or the characteristic impedance $Z^\bullet=\left(Z_1^\bullet/Y_1^\bullet\right)^{1/2}$. The former is preferable, since it removes the ambiguity on the choice of the branch of the complex square root (Fedorov, private communication).

\end{document}